\documentclass[apj,numberedappendix]{emulateapj}
\usepackage{xcolor}
\usepackage{amsmath}
\usepackage[colorlinks=true,citecolor=black,linkcolor=black,urlcolor=blue]{hyperref}

\newcommand{\rmag}{\>^{0.1}{\rm M}_r-5\log h}

\newcommand{\rmd}{{\rm d}}
\newcommand{\msunhh}{\>h^{-2}\rm M_\odot}
\newcommand{\kpch}{\>h^{-1}{\rm {kpc}}}

\usepackage{color}
\definecolor{darkgreen}{rgb}{0.0,0.5,0.0}

\shorttitle{Galaxy-Galaxy lensing in SDSS}
\shortauthors{Luo et al.}

\begin{document}

%%%%%%%%%%%%%%%%%%%%%%%%%%%%%%%%%%%%%%%%%%%%%%%%%%%%%%%%%%%%%%%%%%%%%%%%%%

\title{Galaxy-Galaxy Weak Lensing Measurements from SDSS: I. Image
  Processing and Lensing signals}

\author{Wentao Luo\altaffilmark{1,4}, Xiaohu Yang\altaffilmark{2,3},
  Jun Zhang\altaffilmark{2}, Dylan Tweed\altaffilmark{2}, Liping
  Fu\altaffilmark{5}, H.J. Mo\altaffilmark{6,7}, Frank C. van den
  Bosch\altaffilmark{8}, Chenggang Shu\altaffilmark{5}, Ran
  Li\altaffilmark{9}, Nan Li\altaffilmark{10,11,12}, Xiangkun
  Liu\altaffilmark{13}, Chuzhong Pan\altaffilmark{13}, Yiran
  Wang\altaffilmark{14}, Mario Radovich\altaffilmark{15,16} }

\altaffiltext{1}{ Key Laboratory for Research in Galaxies and
  Cosmology, Shanghai Astronomical Observatory, Nandan Road 80,
  Shanghai 200030, China; E-mail:  \href{mailto:walt@shao.ac.cn}{walt@shao.ac.cn}}

\altaffiltext{2} {Center for Astronomy and Astrophysics, Shanghai Jiao
  Tong University, Shanghai 200240, China; E-mail: \href{mailto:xyang@sjtu.edu.cn}{xyang@sjtu.edu.cn}}
  
\altaffiltext{3} {IFSA Collaborative Innovation Center, Shanghai Jiao
  Tong University, Shanghai 200240, China}

\altaffiltext{4}{ Department of Physics, Carnegie Mellon University,
  Pittsburgh, PA 15213, USA}

\altaffiltext{5} { Shanghai Key Lab for Astrophysics, Shanghai Normal
  University, 100 Guilin Road, 200234, Shanghai, China}

\altaffiltext{6} {Department of Astronomy, University of Massachusetts, Amherst
  MA 01003-9305, USA}

\altaffiltext{7} {Physics Department and Center for Astrophysics, Tsinghua University,
  Beijing 10084, China}

\altaffiltext{8} {Department of Astronomy, Yale University, PO Box 208101,
  New Haven, CT 06520-8101, USA}

\altaffiltext{9} {Key laboratory for Computational Astrophysics, 
  Partner Group of the Max Planck Institute for Astrophysics, 
  National Astronomical Observatories, Chinese Academy of Sciences, 
  Beijing, 100012, China }

\altaffiltext{10} {Department of Astronomy \& Astrophysics, The University of Chicago,
  5640 South Ellis Avenue, Chicago, IL 60637, USA}

\altaffiltext{11} {Argonne National Laboratory, 9700 South Cass Avenue B109,
  Lemont, IL 60439, USA}

\altaffiltext{12} {Kavli Institute for Cosmological Physics at the University of Chicago,
  Chicago, IL 60637, USA}

\altaffiltext{13} {Department of Astronomy, Peking University, Beijing
  100871, China}

\altaffiltext{14} {Department of Astronomy, University of Illinois at
  Urbana-Champaign, 1002 W Green St., Urbana, IL 61801, USA}

\altaffiltext{15} {INAF-Osservatorio Astronomico di Napoli, via Moiariello 16,
  I-80131 Napoli, Italy}

\altaffiltext{16} {INAF-Osservatorio Astronomico di Padova, vicolo
  dell'Osservatorio 5, I-35122 Padova, Italy}

\begin{abstract}
  As the first paper in a series on the study of the galaxy-galaxy
  lensing from Sloan Digital Sky Survey Data Release 7 (SDSS DR7), we
  present our image processing pipeline that corrects the systematics
  primarily introduced by the Point Spread Function (PSF).  Using this
  pipeline, we processed SDSS DR7 imaging data in $r$ band and
  generated a background galaxy catalog containing the shape
  information of each galaxy.  Based on our own shape measurements of
  the galaxy images from SDSS DR7, we extract the galaxy-galaxy (GG)
  lensing signals around foreground spectroscopic galaxies binned in
  different luminosity and stellar mass. The overall signals are in
  good agreement with those obtained by \citet{Mandelbaum2005,
    Mandelbaum2006} from the SDSS DR4.  The results in this paper with
  higher signal to noise ratio is due to the larger survey area than
  SDSS DR4, confirm that more luminous/massive galaxies bear stronger
  GG lensing signal.  We also divide the foreground galaxies into
  red/blue and star forming/quenched subsamples and measured their GG
  lensing signals, respectively.  We find that, at a specific stellar
  mass/luminosity, the red/quenched galaxies have relatively stronger
  GG lensing signals than their counterparts especially at large
  radii. These GG lensing signals can be used to probe the galaxy-halo
  mass relations and their environmental dependences in the halo
  occupation or conditional luminosity function framework.  Our data
  are made publicly available in
  \url{http://gax.shao.ac.cn/wtluo/weak\_lensing/wl\_sdss\_dr7.tar.gz}.
\end{abstract}

%%%%%%%%%%%%%%%%%%%%%%%%%%%%%%%%%%%%%%%%%%%%%%%%%%%%%%

\keywords{(cosmology:) gravitational lensing; galaxies: clusters: general}

\section{Introduction}
\label{sec_intro}

The nature of dark matter remains a mystery in the current paradigm of
structure formation \citep[see][for a review]{Bertone2005}.  Although
many experiments have been proposed to directly detect signatures of
dark matter, such as particle annihilation, particle decay, and
interaction with other particles \citep[see][for a review]{Feng2010},
the main avenue to probe the existence and properties of dark matter
is still through the gravitational potentials associated with the
structures in the dark matter distribution.

One promising way to detect the gravitational effects of dark matter
structures is through their gravitational lensing effect, in which
light rays from distant sources are bent by foreground massive objects
such as galaxies or clusters of galaxies residing in massive dark
matter halos.  In the case of galaxies, the multiple images prediction
was first observationally confirmed by \citet{Walsh1979}. Since then
more and more strong lensing systems were found and analyzed
\citep[e.g.][]{Oguri2002, Kneib2004, Broadhurst2005, Treu2006,
  Cabanac2007, Bolton2008, Coe2013}.  In addition, smaller distortions
in galaxy images have been detected in large surveys, such as SDSS,
CFHTLS, and SUBARU weak lensing surveys. These are referred to as weak
lensing effects and have been studied very extensively in the past
decade \citep{Kaiser1995, Sheldon2004, Mandelbaum2005, Mandelbaum2006,
  Wittman2006, Fu2008, Bernstein2009,Cacciato2009, Oguri2009,
  George2012, Li2013, Mandelbaum2013, Li2014}.

Weak gravitational lensing studies are further sub-divided into two
categories: lensing effects based on individual massive systems, such
as clusters of galaxies, galaxy-galaxy lensing, which relies on the
stacking of lensing signals around many galaxies. For a deep survey
such as CFHTLenS \citep{Heymans2012}, DES \citep{Jarvis2015}, DLS
\citep{Wittman2006}, EUCLID \citep{Refregier2010}, LSST
\citep{LSST2009}, KIDS \citep{Kuijken2015} and SUBARU weak lensing
survey \citep{Kaifu1998, Umetsu2007}, the number density of background
galaxies around a single cluster is sufficient to measure the weak
lensing signals with high S/N ratio, so that the mass and shape of the
dark matter distribution can be obtained \citep{Oguri2010}. For
shallower surveys and for less massive systems, such as SDSS
\citep{York2000}, stacking lensing signals around many systems is the
only way to measure the weak lensing effects with sufficient S/N
ratio. Although unable to give dark matter distributions associated
with individual systems, galaxy-galaxy lensing provides a powerful
tool to estimate the average mass and profile of dark matter halos
around galaxies with certain properties, as the luminosity, stellar
mass, etc.

In principle, weak gravitational lensing can provide a clean
measurement of the total mass distribution of the lens system.
However, the lensing signals are weak and a number of effects need to
be understood and modeled accurately to obtain reliable results. These
include uncertainties in photometric redshifts, intrinsic alignment,
source selection bias and mask effect \citep{Yang2003,Mandelbaum2005,
  Mandelbaum2006, Yang2006a, Mandelbaum2008, Mandelbaum2009a,
  Mandelbaum2009b, Li2009, Sheldon2009, Liu2015}. In addition,
accurate image measurements are absolutely essential in galaxy-galaxy
lensing studies. Thus, for any weak lensing survey, an image
processing pipeline has to be developed first and validated by a
series of test simulations, such as STEP (Shear TEsting Program)
\citep{Heymans2006, Massey2007}, Great08 \citep{Bridle2009}, Great 10
\citep{Kitching2010}, GREAT3 \citep{Mandelbaum2014} or Kaggle -- the
dark matter mapping competition\footnote{Supported by NASA \& the
  Royal Astronomical Society.}. Other independent softwares, such
SHERA \citep[hereafter M12]{Mandelbaum2012}, have also been designed
for specific surveys.

Many groups have developed image processing pipelines devoted to
improving the accuracy of shape measurements for weak lensing studies
\citep{Kaiser1995, Bertin1996, Maoli2000, Rhodes2000, vanWaerbeke2001,
  Bernstein2002,Bridle2002, Refregier2003, Bacon2003, Hirata2003,
  Heymans2005, Zhang2010, Zhang2011,Bernstein2014, Zhang2015}.  Among
these, Lensfit \citep{Miller2007, Miller2013, Kitching2008} applies a
Bayesian based model-fitting approach; BFD (Bayesian Fourier Domain)
method \citep{Bernstein2014} carries out Bayesian analysis in the
Fourier domain, using the distribution of un-lensed galaxy moments as
a prior, and the Fourier\_Quad method developed by \citep{Zhang2010,
  Zhang2011, Zhang2015} uses image moments in the Fourier Domain.

In this paper we attempt to develop an image processing pipeline for
weak lensing studies by combining the \citet[herefater
BJ02]{Bernstein2002} method (see Appendix \ref{bj02_detail} for
details) with the re-Gaussianization method introduced in
\citet[hereafter HS03]{Hirata2003}.  We test the performance of our
pipeline using a number of commonly adopted simulations, and we apply
our method to the SDSS data. The structure of the paper is as follows.
In Section \ref{sec_method}, we describe the procedures used to
construct our image processing pipeline.  The pipeline is tested using
simulations in Section \ref{sec_test}.  Section \ref{sec_application}
presents the application of our pipeline to the SDSS DR7 data, along
with the galaxy-galaxy lensing results obtained for galaxies of
different luminosities and colors. Finally, we summarize our results
in Section \ref{sec_summary}.  In addition, some details of our method
are given in Appendix~\ref{bj02_detail}, some tests on systematic
errors are made in Appendix \ref{sec_systematic}, and our main results
for the SDSS data are listed inTables presented in Appendix
\ref{sec_ESD}.  All the galaxy-galaxy lensing data shown in this paper
can be downloaded from
\href{http://gax.shao.ac.cn/wtluo/weak\_lensing/wl\_sdss\_dr7.tar.gz}
{http://gax.shao.ac.cn/wtluo/weak\_lensing/wl\_sdss\_dr7.tar.gz}.

\section{Image processing pipeline}
\label{sec_method}

The goal of our pipeline is to measure, for each observed galaxy image
$I_{\rm obs}(\mathbf{x})$, two ellipticity parameters $e_1$ and $e_2$
(to be defined below) that describe the intrinsic shape of the
galaxy. However, the observed image is the convolution between the
intrinsic galaxy image $I_{int}(\mathbf{x})$ and the PSF
$P(\mathbf{x})$,
\begin{equation} 
I_{\rm obs}(\mathbf{x})=I_{\rm
        int}(\mathbf{x})\otimes P(\mathbf{x})\,,  
\end{equation} 
where $I_{\rm int}(\mathbf{x})$ stands for the intrinsic galaxy surface
brightness and $P(\mathbf{x})$ is the PSF.  Formally, the impact of 
the PSF on the ellipticity parameters can be written as  
\begin{equation} 
e_i^{\rm obs}=(1+m)e_i^{\rm int}+c\,, 
\end{equation} 
where $i=1,2$. PSF anisotropy causes a non-zero additive error $c$,
while PSF smearing causes a non-zero multiplicative error $m$.  The
challenge is to develop a reduction pipeline that minimize both $|m|$
and $|c|$.  
Our pipeline consists of the following steps:
\begin{itemize} 
\item Create a kernel function $K(\mathbf{x})$ to correct for the
  PSF anisotropy (see  \S \ref{psf_ani} for details). 
\item Convolve both $I_{\rm obs}(\mathbf{x})$ and $P(\mathbf{x})$ with
  the kernel function $K(\mathbf{x})$, so that we have
  $I_1(\mathbf{x})=I_{\rm obs}(\mathbf{x})\otimes K(\mathbf{x})$
  and $P_1(\mathbf{x})=P(\mathbf{x})\otimes K(\mathbf{x})$. 
\item Measure the sizes $T_I$ and $T_P$, as well as the ellipticity
  parameters $e_1$ and $e_2$, from the surface brightness weighted
  second moments of $I_1$ and $P_1$, $M_{I_1}$ and $M_{P_1}$,
  respectively, using the adaptive Gaussian kernel method as described
  in \S \ref{shape_params}. 
\item Re-Gaussianize $P_1$ and $I_1$ using the method of HS03. This
  results in $P_{\rm RG}=G(M_{P_1})$ and $I_{\rm RG}=
  I_1-G(M_{I_1})\otimes[P_1-G(M_{P_1})])$. Here $G(M_{P_1})$ and
  $G(M_{I_1})$ are 2D Gaussian functions reconstructed from the same
  second moments $M_{P_1}$ and $M_{I_1}$ obtained from $P_1$ and
  $I_1$.  
\item Expand $I_{\rm RG}$ and $P_{\rm RG}$ in terms of the Quantum
  Harmonic Oscillator (QHO) eigenfunctions, as described in \S
  \ref{psf_ani}. Compute $\beta^{I}_{22}=b^{I}_{22}/b^{I}_{00}$ and
  $\beta^{P}_{22}=b^{P}_{22}/b^{P}_{00}$, where $b^{I}_{ij}$ and
  $b^{P}_{ij}$ are the coefficients of the QHO expansions of $I_{\rm RG}$ and
  $P_{\rm RG}$, respectively.  
 \item Calculate the resolution factor $\cal R$, which is defined as,
  \begin{equation} {\cal
      R}=1-\frac{T_P(1-\beta_{22}^p)/(1+\beta_{22}^p)}
    {T_I(1-\beta_{22}^I)/(1+\beta_{22}^I)}\,,
\label{eq:eq_R}
\end{equation}
and correct the ellipticity parameters according to  
$e_1^{\rm corr}=e_1/{\cal R}$ and $e_2^{\rm corr}=e_2/{\cal R}$. 
\item Rotate $(e_1^{\rm corr},e_2^{\rm corr})$, which are measured
  with respect to the coordinates of the CCD image (i.e., $e_1$ is
  measured in the direction of CCD pixel rows), such that $e_1$ is
  aligned with the direction of increasing right ascension.  This is
  achieved by  the transformation:
\begin{equation}
\left(
      \begin{array}{c}
               e_1^{\rm rot}\\
               e_2^{\rm rot}\\
       \end{array}
\right)=
\left(
      \begin{array}{cc}
               \cos 2\phi & -\sin 2\phi\\
               \sin 2\phi & \cos 2\phi\\
      \end{array}
\right)
\left(
\begin{array}{c}
               e_1^{\rm corr}\\
               e_2^{\rm corr}\\
       \end{array}
\right)\,,
\end{equation}
where $\phi$ is the angle between the North and the direction of the CCD
columns, and is provided in the header of each CCD image.
\item Background noise of each galaxy is estimated from both the sky 
  and the
  dark current as in \citet[hereafter M05]{Mandelbaum2005},
  \begin{equation}\label{eq:sky}
\sigma_{{\rm sky}} =\frac{\sigma^{I}}{{\cal R}F}\sqrt{4\pi n}
\end{equation}
where $\sigma^{I}$ is the size of the galaxy in pixels, $F$ is the
flux and $n$ is the sky and dark current brightness in photons per
pixel.
\end{itemize}

The end result of our pipeline is a catalog listing for each source
image the values, $e_1^{\rm rot}$, $e_2^{\rm rot}$, $\cal R$,
$\sigma_{{\rm sky}}$, $\alpha$, $\delta$, $z_{\rm photo}$, where
$\alpha$, $\delta$, $z_{\rm photo}$ are the RA, DEC and photometric
redshift of the source (galaxy).  Note that we explicitly list $\cal R$
since it is a common practice in galaxy-galaxy lensing
measurements to only select images with $\cal R$ exceeding some 
limiting value. Throughout this paper we follow M05 and only use images 
with ${\cal R}>1/3$.

\subsection{PSF anisotropy correction}
\label{psf_ani}

There are two systematics associated with the PSF.  One is an
isotropic smearing of the original image and the other is an
anisotropic effect which introduces extra shape distortion.  Our image
processing pipeline is designed to correct for both effects.  More
specifically, we use the rounding kernel method of BJ02 for the
anisotropic correction and the re-Gaussianization method of HS03 for
the isotropic correction.  The reason for this combination is that,
according to our test with STEP2 data, the multiplicative error it
produces is the smallest among the other methods (e.g. BJ02 method
alone, re-Gaussianization method alone, and the KSB method).  In this
subsection, we focus on the PSF anisotropy correction.

The basic idea of the rounding kernel method of BJ02 for PSF
anisotropy corrections is to convolve the PSF with a reconstructed
kernel. In the ideal case, the Fourier transformation of the kernel
$K$ is related to the PSF as $\tilde{K}=1/\tilde{P}$ so that the
convolution of $K$ and $P$ in real space is a delta function. In that
case we have:
\begin{equation}
 K(\mathbf{x})\otimes P(\mathbf{x})=\delta(\mathbf{x})\,,
\end{equation}
and
\begin{equation}
 I_{\rm int}(\mathbf{x})=I_{\rm obs}(\mathbf{x})\otimes K(\mathbf{x}) \,,
\end{equation}
where $I_{\rm obs}$ is the observed image, $I_{\rm int}$ the intrinsic
image, and $ K(\mathbf{x})$ represents the reconstructed kernel [see
Eqs.\,(7.1) - (7.4) in BJ02].  In real applications, the PSF is not
modeled perfectly, a better kernel approximation is needed to serve
our purpose.  To this end, we expand the PSF with the Quantum Harmonic
Oscillator (QHO) eigenfunctions,
 \begin{equation}\label{eq4}
P=\sum_{p,q}b_{pq}\phi_{pq}^{\sigma}(r,\theta),
\end{equation}
and write its convolution with $K$ as 
\begin{equation}\label{eq4xx}
K\otimes P=\sum_{p,q}\mathrm{b}_{pq}^{*}\phi_{pq}^{\sigma}(r,\theta), 
\end{equation}
where 
\begin{equation}
  \phi_{pq}^{\sigma}(r,\theta)=\frac{(-1)^q}{\sqrt{\pi}\sigma^{2}}
  \sqrt{\frac{q!}{p!}}(r/\sigma)^{m}e^{im\theta}e^{-r^2/2\sigma^{2}}L_q^{m}(r^{2}/\sigma^{2})\,.
\end{equation}
$L_q^{m}$ are the Laguerre polynomials obeying $m=p-q$ and $\sigma$ is
the size of the object in pixel units.  If $\mathrm{b}_{pq}^*$
satisfies
\begin{equation}
\mathrm{b}_{pq}^*=[(-1)^p/\sqrt{\pi}]\delta_{pq}
\end{equation}
up-to some order $N=p+q$, the PSF anisotropy is then ideally removed
by the reconstructed kernel. Note that 
$\mathrm{b}_{10}^*$ can be set to 0 if the PSF centroid is properly measured,
and the dominant bias is introduced by $\mathrm{b}_{20}^*$,
$\mathrm{b}_{31}^*$ and so on (see Appendix \ref{bj02_detail} for 
details).

\subsection{Shape parameters}
\label{shape_params}

In this subsection, we outline a few parameters that are important in
galaxy-galaxy lensing shear measurements as well as in our image
processing pipeline.  The shape parameters, $e_1$ and $e_2$, are
obtained from the surface brightness-weighted second moment of the
2-dimensional galaxy image \citep{Kaiser1995}, 
\begin{equation}
  M_{ij}=\frac{\sum G(\mathbf{x})
    I(\mathbf{x})(\mathbf{x-x_0})_{i}(\mathbf{x-x_0})_{j}}
  {\sum G(\mathbf{x})I(\mathbf{x})}\,
\end{equation}   
where $i,j=x,y$ and $I(\mathbf{x})$ is the surface brightness at the
pixel located at $\mathbf{x}$. The function $G(\mathbf{x})$ is an adaptive
Gaussian kernel (see section 2.1 in HS03) used to avoid divergent noise
\citep{Kaiser1995}:
\begin{equation}
G(\mathbf{x})=\exp\left[-0.5*(\mathbf{x-x_0})^{\rm T}M^{-1}(\mathbf{x-x_0})\right],
\end{equation}
with $\mathbf{x_0}$ being the centroid vector:
\begin{equation}
  \mathbf{x_0}=\frac{\int
    \mathbf{x}G(\mathbf{x})I(\mathbf{x})d^2\mathbf{x}}
  {\int G(\mathbf{x})I(\mathbf{x})d^2\mathbf{x}}\,.
\end{equation}

Following convention, the ellipticity parameters $e_1$ and $e_2$ are
respectively defined as the compressions along a fiducial direction
(e.g. $x$) and along a direction rotated 45 degrees with respect to
it. The size $T$ is defined as the trace of the moment tensor.
Thus, 
\begin{eqnarray}
e_1&=&\frac{M_{xx}-M_{yy}}{M_{xx}+M_{yy}}\nonumber\\
e_2&=&\frac{2M_{xy}}{M_{xx}+M_{yy}}\nonumber \\
T&=&M_{xx}+M_{yy}.
\end{eqnarray}

\subsection{Re-Gaussianization}
\label{regauss}

PSF isotropic effect a.k.a smearing effect, dilutes the value of
ellipticity and therefore leads to shear underestimation. The method
we adopt here to correct this effect is the re-Gaussianization method
from HS03. It consists of applying a resolution factor ${\cal R}$
(Eq. \ref{eq:eq_R}) to correct for the ellipticity parameters. Note,
however, as pointed out in HS03, Eq. \ref{eq:eq_R} (Eq. 14 in HS03) is
only valid when both PSF and galaxy images are Gaussian, which does
not apply to real observations.  HS03 reconstructed a Gaussian PSF
model using the second moments from the PSF, and then corrected the
galaxy image for the effect of the residuals. We assume that,
Eq. \ref{eq:eq_R} is valid after these treatments. The related
processes are called re-Gaussianization and carried out as follows.

We first construct a Gaussian PSF from the real PSF $P(\mathbf{x})$
using the second moment covariance matrix $M_P$,
\begin{equation}
  G(\mathbf{x})=\frac{1}{2\pi\sqrt{\det M_P}}
  \exp\left(-\frac{1}{2}\mathbf{x}^{\rm T}M_P^{-1}\mathbf{x}\right)\,,
\end{equation}
with the residual being,
\begin{equation}
  \epsilon(\mathbf{x})=P(\mathbf{x})-G(\mathbf{x})\,.
\end{equation}
The galaxy image $I_{\rm obs}(\mathbf{x})$ then satisfies
\begin{equation}
  I_{\rm obs}(\mathbf{x})=P(\mathbf{x})\otimes I_{\rm int}(\mathbf{x})=
  G(\mathbf{x})\otimes  I_{\rm int}(\mathbf{x})
  +\epsilon(\mathbf{x})\otimes  I_{\rm int}(\mathbf{x})\,,
\end{equation}
where $ I_{\rm int}(\mathbf{x})$ is the intrinsic brightness
distribution of the galaxy. 

Next, we approximate the galaxy image also with a Gaussian
distribution from its second moments matrix, $M'_I$,
\begin{equation}
  I^0_{\rm obs}(\mathbf{x})=\frac{1}{2\pi\sqrt{\det M'_I}}
  \exp\left(-\frac{1}{2}\mathbf{x}^{\rm T}{M'_I}^{-1}\mathbf{x}\right)\,,
\end{equation}
where $M'_I=M_I-M_P$ denotes the galaxy second moments matrix once the
PSF contribution has been subtracted.  

Finally, an image, corrected for the residual between real PSF and
Gaussian PSF, is obtained using
\begin{equation}
  I'_{\rm obs}(\mathbf{x})=I_{\rm obs}(\mathbf{x})-
  \epsilon (\mathbf{x})\otimes  I^0_{\rm obs}(\mathbf{x})\,.
\end{equation}
In our pipeline, we compute the $\beta_{22}$ from 
$I'_{\rm obs}(\mathbf{x})$ and $G(\mathbf{x})$.

\subsection{Shear estimator}
\label{shear_est}

Once we have processed all the source images, we can obtain the shear
signals $\gamma$ along any desired directions. 
We first compute the responsiveness $\bar{R}$ of our survey
galaxies which is defined as
\begin{equation}
\bar{R}\equiv 1- {1\over N} \sum_{i=1}^{N} (e_1^{\rm rot})^2 \,,
\end{equation}
where $N$ is the total number of source images with ${\cal R}>1/3$.
Next we compute the shear components $\gamma_1$ and $\gamma_2$ using 
all the images sampling the local shear field:
\begin{equation}
\gamma_l=\frac{1}{2\bar{R}}\frac{\sum w_ie_l^{\rm rot}}{\sum w_i}\,,
\end{equation}
where $l=1,2$ and $w_i$ is a weighting function. Each source image is
weighted by 
\begin{equation}
  w=\frac{1}{\sigma_{\rm sky}^2+\sigma_{\rm shape}^2}\,,
\end{equation}
where $\sigma_{\rm sky}$ is the background noise estimated using
Eq. \ref{eq:sky} and $\sigma_{\rm shape}$ is the shape noise.  For a
sample of background galaxies, the shape noise is defined as the
variance of their ellipticities.

Observationally, the tangential shear $\gamma_{\rm T}$ as a function 
of radius around foreground lens galaxies is estimated as 
 \begin{equation}
\gamma_{\rm T}(R)=\frac{1}{2\bar{R}}\frac{\sum w_ie_{\rm T}}{\sum w_i}\,,
\end{equation}
where $e_{\rm T}$ is given by 
\begin{equation}
\left(
      \begin{array}{c}
               e_{\rm T}\\
               e_{45^0}\\
       \end{array}
\right)=
\left(
      \begin{array}{cc}
               \cos 2\theta & -\sin 2\theta\\
               \sin 2\theta & \cos 2\theta\\
      \end{array}
\right)
\left(
\begin{array}{c}
               e_1^{\rm rot}\\
               e_2^{\rm rot}\\
       \end{array}
\right)\,,
\end{equation}
with $\theta$ the angle between the line connecting the lens and the
source and the direction of increasing right ascension. So defined,
$e_{\rm T}$ is the shape parameter along the tangential direction
around the lens.

\section{Testing the pipeline with simulations}
\label{sec_test}

Before applying our pipeline to real data, we benchmark test it using
simulated images. These contain input shear signals as well as
observational effects, such as PSF, sky background noise and
pixellization. The two simulations catalogs used here are SHERA (SHEar
Reconvolution Analysis) developed by M12, and GREAT3 as described in
\citep{Mandelbaum2014}.

\subsection{Testing with SHERA}

\subsubsection{SHERA data}

\begin{figure*}
\centering
\includegraphics[width=5cm,height=5cm]{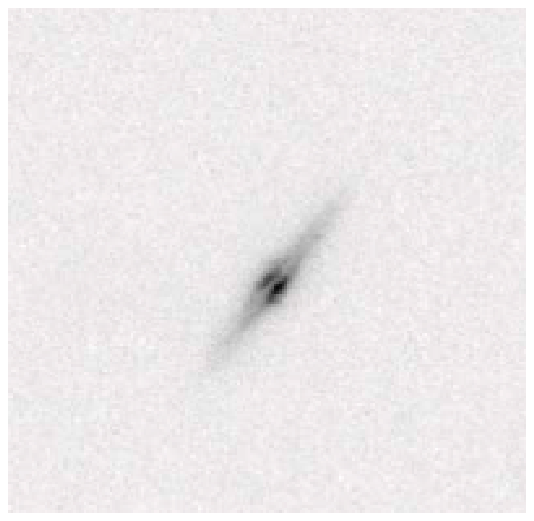}
\includegraphics[width=5cm,height=5cm]{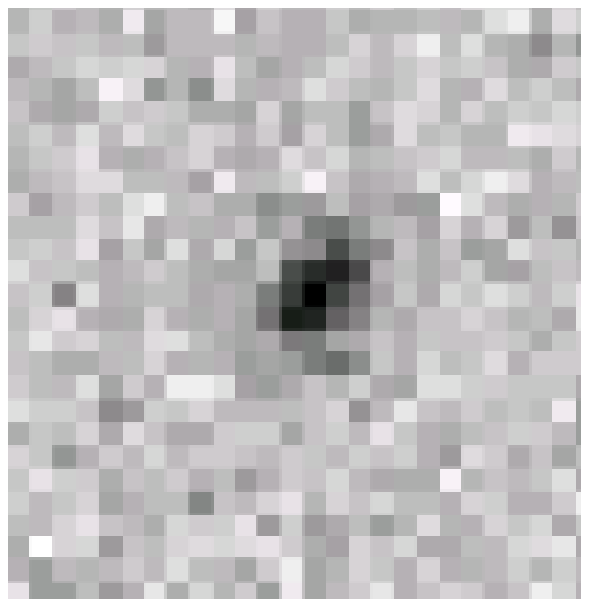}
\includegraphics[width=5cm,height=5cm]{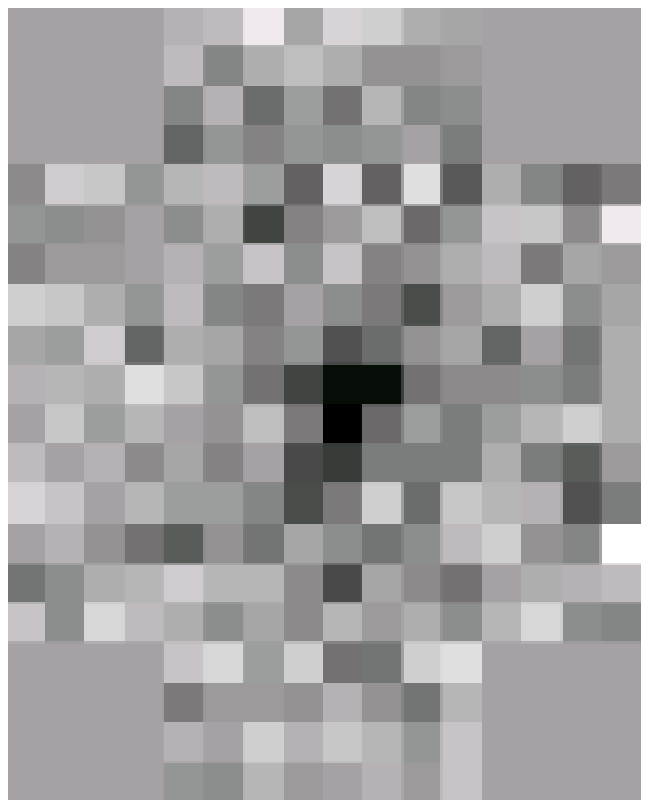}
\caption{COSMOS image (left panel),
  SHERA simulated image (middle panel) and SDSS real image of the
  same galaxy (right panel).}
  \label{fig:image}
\end{figure*}

SHERA (M12) is designed to test the accuracy of shape measurement
pipelines for ground-based images. It uses Cosmological Evolution
Survey (COSMOS) images as input. The output are low-resolution images
expected from a given ground-based observation. Parameters such as
pixel size, PSF size, and sky background, are set in accordance to
SDSS data.  The weak lensing shear signal is added to each image using
the following equation:
\begin{equation}
\left(
      \begin{array}{c}
               x^u\\
               y^u\\
       \end{array}
\right)=
\left(
      \begin{array}{cc}
               1-\kappa-\gamma_1 & -\gamma_2\\
               -\gamma_2 & 1-\kappa+\gamma_1\\
      \end{array}
\right)
\left(
\begin{array}{c}
               x^l\\
               y^l\\
       \end{array}
\right)\,,
\end{equation}
where $(x^u,y^u)$ are the un-lensed coordinates, and $(x^l,y^l)$ the lensed
ones. The input shear $(\gamma_1, \gamma_2)$ 
are randomly generated ranging from $-0.05$ to $0.05$.

The input galaxy image catalog is constructed from COSMOS ACS field
\citep{Koekemoer2007,Scoville2007a, Scoville2007b} following the
method described in \citet{Leauthaud2007}. The survey field is a
$1.64$ square degree region centered at 10:00:28.6$+$0.l2:12:21.0
(J2000).  The images are corrected for charge transfer inefficiency
\citep[CTI]{Massey2010}, geometric distortion, sky subtraction and
cosmic rays, and are further dithered using multi-drizzle algorithm.
The final production is a co-added image of $7000\times 7000$ pixels
with a scale of 0.03"/pixel.  Further cuts are applied to fulfill the
special requirements of SHERA, as described in $\S 4.1$ of M12.

With the above criteria, 30,225 galaxies are selected.  To mimic the
SDSS images, additional galaxies are discarded either because these
sources are undetectable in SDSS or because their sizes are smaller
than the SDSS PSF, as detailed in M12. The final sample contains
26,113 galaxies.

\subsubsection{PSF matching}

The high resolution images obtained above are transformed into low
resolution ones by PSF matching, i.e. by first de-convolving the 
images with the space PSF and then convolving them with the 
ground-based PSF.  In Fourier space, this is mathematically given by 
\begin{equation}
\tilde{I^g}=\frac{\tilde{G^g}}{\tilde{G^s}}\tilde{I^s}\,,
\end{equation}
where $I^g$ and $G^g$ are the ground-based brightness distribution and
PSF, respectively, whereas $I^s$ and $G^s$ are the corresponding
space-based quantities. This PSF matching works as long as the power
spectrum of the space PSF is larger than the one of the ground PSF for
all $k$; otherwise it leads to ringing effect in the new image.  As
shown in Fig.\,2 of M12, the power spectrum of SDSS PSF is smaller
than the one of COSMOS at all wave numbers, and so the PSF matching
can be done safely.

In addition to the PSF, the noise level at the position of COSMOS in
the SDSS imaging should also be taken into account.
Fig.~\ref{fig:image} shows the COSMOS image of a typical disk galaxy
(left), the SHERA simulated SDSS image (middle) and the real SDSS
fpAtlas image (right) of the same galaxy. The bulge and disk
components can be clearly identified from the original COSMOS image,
whereas in SDSS only a small number of pixels brighter than the
detection limit (22.0 in r band) can be identified.  We downgrade the
high resolution COSMOS images to low resolution SDSS images. During
this process, we miss 2.2 percent of the objects because of masking,
which leaves a total of 25,527 images.

\subsubsection{SHERA testing results}

\begin{figure}
\centering
\includegraphics[width=8cm]{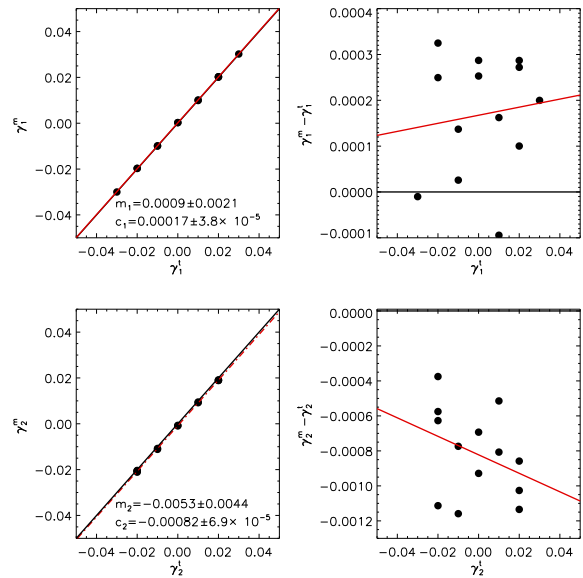}
\caption{Correlation between the true (input) and measured shear 
components $\gamma_1$ (upper left) and $\gamma_2$ 
(lower left). The corresponding residuals  are plotted 
against the input values  in the right panels. The red lines are linear fits.}
\label{fig:shera}
\end{figure}

Using the mock SDSS images obtained above, we follow M12 and rotate
each image 90 degree in order to eliminate the effect from intrinsic
galaxy shape. In the spirit of making a fair comparisons with the results
of M12, the sky background and Poisson noise are not added to the
simulation, so as to assess the performance of the PSF correction alone.
Due to the size cut, only about 11,700 (44\%) galaxies are selected
for the final shear measurements.

We measured the two shear components $\gamma_1^{measure}$ and
$\gamma_2^{measure}$, and compare them to the input signals in
Fig. \ref{fig:shera}.  The upper-left and low-left panels are the
one-to-one correlations for the two components, while the right panels
are the corresponding residuals plotted against the input signals.
The red lines are the linear fit to the data points.  We use the
standard terminology of multiplicative error (including PSF smearing
effect and other unknown bias in the measurement method itself) and
additive error (mostly from PSF anisotropy) to relate the input signal
and the measured signal:
\begin{gather}
\gamma_i^{measure}=(1+m_i)\gamma_i^{input}+c_i
~~~~(i=1,2)
\end{gather}
where $m_i$ and $c_i$ represent the two types of errors.  In general,
our pipeline achieves $<1\%$ in the multiplicative error, with
$m_1=0.09\% \pm 0.0021$ and $m_2=0.53\%\pm 0.0044$, and $<0.1\%$ in
the additive error, with $c_1= 0.00017\pm 3.8\times 10^{-5}$ and $c_2=
0.00082\pm 6.9\times 10^{-5}$.  The fact that the multiplicative error
in $\gamma_2$ is larger than the one in $\gamma_1$ is due to
pixellization.  In \citet{Mandelbaum2012} the corresponding
multiplicative errors are $m_1=-1.6\% \pm 0.001$ and $m_2=-2.7\% \pm
0.001$, and the additive errors are $c_1= 0.00028\pm 1.0\times
10^{-5}$ and $c_2 = -0.00011\pm 1.0\times 10^{-5}$.  These values
demonstrate that the performances of the two pipelines are favorably
comparable.

However, some shortcomings of our pipeline appear during our tests.
When strong sky background and Poisson noise are added, our pipeline
sometimes suffers from non-convergence either during the calculation
of the adaptive moments or during the estimation of the coefficients
(see Eq. \ref{eq:kernel} in Appendix \ref{bj02_detail}).  Thus, our
pipeline cannot provide shape measurements for images with too low
qualities.  This reduces the number of sources that can be used for
lensing studies. Due to the fact that the COSMOS image sample is
small, we do not perform further tests with noise as in M12. The
convergence problem is also not discussed further, because it is
difficult to determine whether it is caused by the iteration of
adaptive moments or by the procedure constraining the $k_{ij}$ (see
Appendix \ref{bj02_detail}).  Nevertheless, as we will show in next
subsection, even with the reduced number of sources, our pipeline
provides lensing signals that are competitive to other methods or
implementations.

\subsection{GREAT3 }

GREAT3 (GRavitational lEnsing Accuracy Test 03) \citep{Mandelbaum2014}
is the continuation of the testing projects STEP \citep[Shear TEsting
Program,][]{Heymans2005}, STEP2 \citep{Massey2007}, GREAT08
\citep{Bridle2009} and GREAT10 \citep{Kitching2010}. All of these code
comparison projects are designed to compare the performances of
different shape measurement methods in different observational
conditions. From STEP to GREAT3, different PSFs, pixel sizes, galaxy
morphologies are adopted. In particular, GREAT3 uses controlled galaxy
morphologies generated with Shapeless \citep{Refregier2003}, real
galaxy morphologies obtained from COSMOS, co-added multiply observed
images, variable PSF, and variable shears. Five major branches of
simulations are generated using GalSim \citep{Rowe2015}: (i) a
controlled sample generated with parametric (single or double
S\'ersic) galaxy models; (ii) real galaxy sample with realistic
morphology from HST COMOS dataset; (iii) multiple-epoch sample
containing six images combined by dithering; (iv) sample with variable
PSF that is reconstructed from star images; (v) a sample that includes
all the above procedures.  Each major branch is further divided into
ground versus space, and constant versus variable shear sub-branches.

For the constant shear datasets, 10,000 galaxies with shear are
simulated. In order to cancel the effect of galaxy intrinsic shape,
GREAT3 applies the same rotation method as in the STEP2 simulation
\citep{Massey2007}. The basic idea is to use the fact that the shape
is a spin-two quantity, meaning that the sum of the original and
90-degree rotated ellipticity is zero.

Our pipeline participated in the controlled ground constant, the
controlled space constant, the real ground constant and the real space
constant tests. We labelled our implementation as BJ02+HS03 within
this project.  Overall, it ranks 15 among a total of 26 participating
pipelines.  As mentioned earlier, our pipeline suffers from a
non-convergence problem. Together with the size cut using the
resolution factor in Eq.~(\ref{eq:eq_R}), only about 40\% galaxies are
used in the competition.  Among our submissions, we found that the
best weighting scheme for our pipeline was to take the inverse of the
shape noise and errors from ellipticity as in \citet[herefater
M15]{Mandelbaum2015}.  The more detailed information and results about
the GREAT3 competition can be found in M15.

\section{Application to the SDSS DR7}
\label{sec_application}

Since our pipeline proved to be reliable, we processed the SDSS DR7
\citep{Abazajian2009} $r$ band imaging data.  The SDSS
\citep{York2000} consists of three imaging and spectroscopic surveys
(Legacy, SEGUE, and Supernova), using a 2.5m telescope at Apache Point
Observatory in Southern New Mexico. The SDSS photometric camera has
two TDI (Time-Delay-and-Integrate) CCD scanning arrays
\citep{Gunn1998}.  One is a $6\times 5$ CCD array, with each of the
CCD having $2048\times 2048$ pixels (24 $\mu m \approx 3 arcseconds$
on the sky) for five-band photometry, and the other is a 24
$2048\times 400$ CCD array used for astrometry and focus
monitoring. The DR7 imaging data, with \textit{u}, \textit{g},
\textit{r}, \textit{i} and \textit{z} band, covers about 8423 square
degrees of the LEGACY sky ($\sim$230 million distinct photometric
objects) and about 3240 square degrees of SEGUE sky, ($\sim$127
million distinct objects, including many stars at low latitude). The
total number of objects identified as galaxies is around 150 million.

\begin{figure}
\centering
\includegraphics[width=9cm,height=9cm]{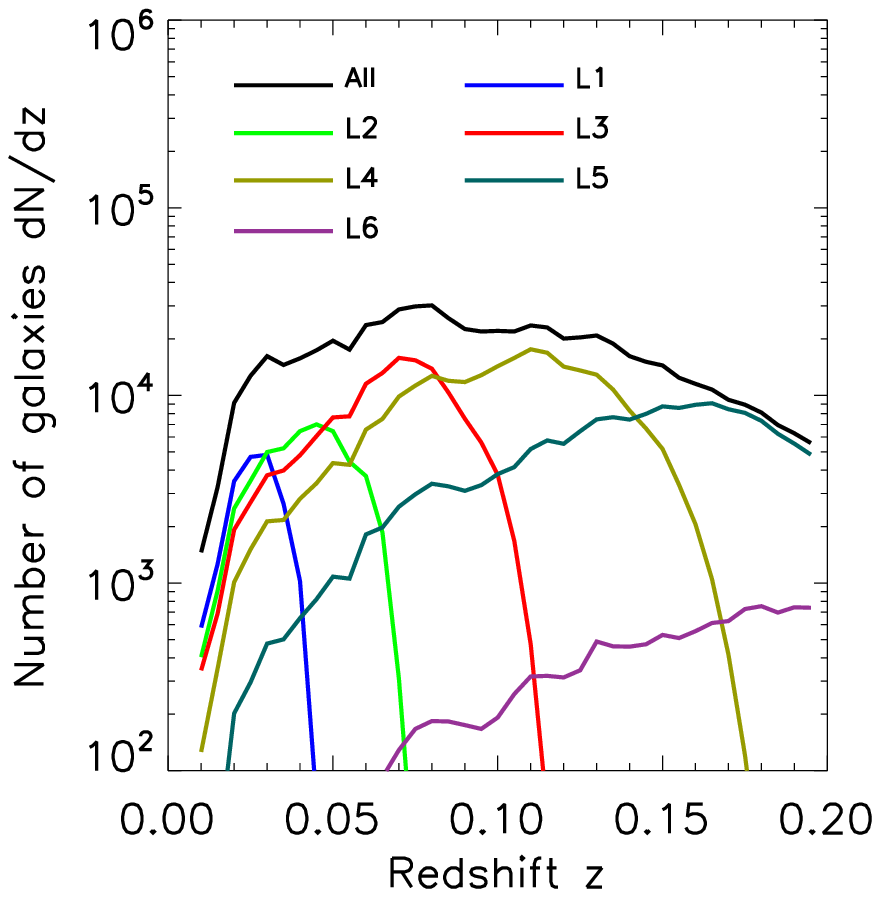}
\includegraphics[width=9cm,height=9cm]{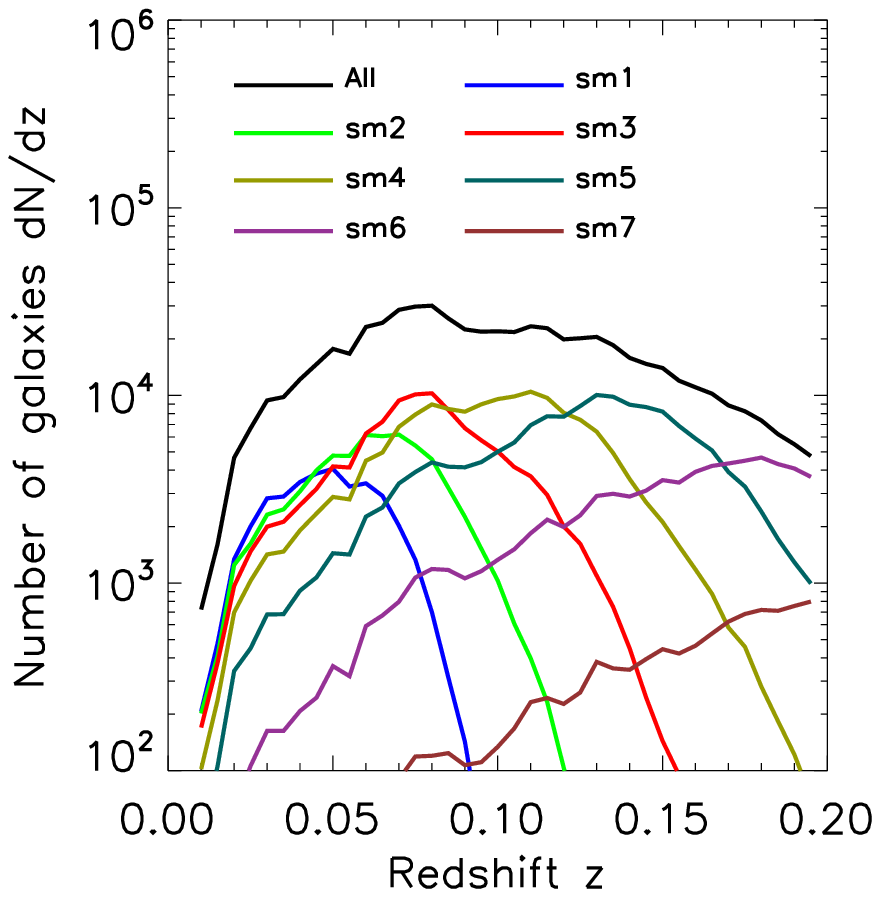}
\caption{The redshift distribution of lens samples binned in
  luminosity (upper panel), and in stellar mass (lower panel).}
  \label{fig:zlens}
\end{figure}

\begin{table}[h!]
\begin{center}
  \caption{\label{tab:tbl-1} Properties of the six lens samples
    created for this paper. We indicate the number of galaxies in the
    equivalent samples in \citet{Mandelbaum2005} $N_{M05}$, to be
    compared to our number $N_{gal}$}.
\begin{tabular}{ccccccc}
\hline
Sample & $M_r$ & $N_{gal}$ & $N_{M05}$ & $\langle z \rangle $ & $\sigma(z)$ & $\langle L \rangle/L_*$\\
\hline
L1  & $(-18,-17]$  & 18 614  & 6 524   & 0.029 & 0.007  & 0.071 \\
L2  & $(-19,-18]$  & 47 795  & 19 192  & 0.044 & 0.012  & 0.181 \\
L3  & $(-20,-19]$  & 138 988 & 58 848  & 0.069 & 0.020  & 0.450 \\
L4  & $(-21,-20]$  & 249 906 & 104 752 & 0.103 & 0.030  & 1.082 \\
L5  & $(-22,-21]$  & 164 653 & 63 794  & 0.140 & 0.038  & 2.364 \\
L6  & $(-23,-22]$  & 11 453  & 6 499   & 0.150 & 0.037  & 5.146 \\
\hline
\end{tabular}
\end{center}
\end{table}

\begin{table}[h!]
\begin{center}
 \caption{\label{tab:tbl-2} 
   Properties of the 12 lens subsamples obtained from the one in
   Table~\ref{tab:tbl-1} after divided by color.}
\begin{tabular}{ccccc}
\hline
Sample & $N_{gal}$ & $\langle z \rangle $ & $\sigma(z)$ & $\langle L \rangle/L_*$\\
\hline
L1R   & 5 383   & 0.030  & 0.007  & 0.073 \\
L1B   & 13 231  & 0.029  & 0.007  & 0.071 \\
L2R   & 17 471  & 0.045  & 0.013  & 0.186 \\
L2B   & 30 324  & 0.044  & 0.012  & 0.179 \\
L3R   & 67 058  & 0.069  & 0.019  & 0.459 \\
L3B   & 71 930  & 0.069  & 0.019  & 0.443 \\
L4R   & 138 316 & 0.102  & 0.030  & 1.092 \\
L4B   & 111 590 & 0.104  & 0.030  & 1.072 \\
L5R   & 98 808  & 0.141  & 0.038  & 2.378 \\
L5B   & 65 845  & 0.138  & 0.038  & 2.347 \\
L6R   & 6 880   & 0.155  & 0.034  & 5.130 \\
L6B   & 4 573   & 0.141  & 0.037  & 5.182 \\
\hline
\end{tabular}
\end{center}
\end{table}

\begin{table}[h!]
\begin{center}
 \caption{\label{tab:tb2-1} Properties of our seven lens samples
   binned by stellar mass. We compare the number of
   galaxies we use $N_{gal}$ to the one in
    \citet{Mandelbaum2006}.  $M_*$ is in units of
    $\msunhh$. }.
\begin{tabular}{ccccccc}
\hline
Sample & $\log(M_*)$ & $N_{gal}$ & $N_{M06}$ & $\langle z \rangle $ & $\sigma(z)$ & $\langle \log(M_*) \rangle$\\
\hline
sm1  & $[9.38,9.69]$    & 35 269  & 23 474   & 0.029 & 0.007  & 9.55 \\
sm2  & $[9.69,9.99]$    & 62 742  & 40 952   & 0.044 & 0.012  & 9.85 \\
sm3  & $[9.99,10.29]$   & 107 707 & 66 503   & 0.069 & 0.020  & 10.15 \\
sm4  & $[10.29,10.59]$  & 153 787 & 90 019   & 0.103 & 0.030  & 10.45 \\
sm5  & $[10.59,10.89]$  & 155 242 & 82 734   & 0.140 & 0.038  & 10.73 \\
sm6  & $[10.89,11.20]$  & 73 048  & 39 729   & 0.150 & 0.037  & 11.01 \\
sm7  & $[11.20,11.50]$  & 9 807   & 8 096    & 0.150 & 0.037  & 11.29 \\
\hline
\end{tabular}
\end{center}
\end{table}

\begin{table}[h!]
\begin{center}
  \caption{\label{tab:tb2-2}  Sub-samples binned
    in stellar mass and split into red and blue. $M_*$ is in units of
    $\msunhh$.}
\begin{tabular}{ccccc}
\hline
Sample & $N_{gal}$ & $\langle z \rangle $ & $\sigma(z)$ & $\langle \log(M_*) \rangle$\\
\hline
sm1r   & 7 447   & 0.038  & 0.010  & 9.56 \\
sm1b   & 27 522  & 0.054  & 0.016  & 9.55 \\
sm2r   & 19 604  & 0.051  & 0.015  & 9.87 \\
sm2b   & 43 138  & 0.070  & 0.020  & 9.85 \\
sm3r   & 48 669  & 0.069  & 0.019  & 10.16 \\
sm3b   & 59 038  & 0.089  & 0.026  & 10.15 \\
sm4r   & 85 839  & 0.090  & 0.025  & 10.45 \\
sm4b   & 67 948  & 0.113  & 0.032  & 10.44 \\
sm5r   & 102 360 & 0.120  & 0.036  & 10.74 \\
sm5b   & 52 882  & 0.136  & 0.037  & 10.72 \\
sm6r   & 57 063  & 0.149  & 0.037  & 11.01 \\
sm6b   & 15 985  & 0.146  & 0.038  & 10.99 \\
sm7r   & 8224    & 0.158  & 0.034  & 11.29 \\
\hline
\end{tabular}
\end{center}
\end{table}

\begin{table}[h!]
\begin{center}
  \caption{\label{tab:tb2-3} Sub-samples binned
    in stellar mass and split into star-forming galaxies and quenched
    galaxies. $M_*$ is in units of
    $\msunhh$.}
\begin{tabular}{lcccc}
\hline
Sample & $N_{gal}$ & $\langle z \rangle $ & $\sigma(z)$ & $\langle \log(M_*) \rangle$\\
\hline
sm1sf   & 29 460  & 0.053  & 0.016  & 9.55   \\
sm1qu   & 5 809   & 0.038  & 0.011  & 9.56   \\
sm2sf   & 46 544  & 0.068  & 0.020  & 9.85   \\
sm2qu   & 16 198  & 0.051  & 0.016  & 9.87   \\
sm3sf   & 66 138  & 0.086  & 0.027  & 10.15  \\
sm3qu   & 41 569  & 0.069  & 0.019  & 10.16  \\
sm4sf   & 73 606  & 0.109  & 0.031  & 10.44  \\
sm4qu   & 80 181  & 0.090  & 0.026  & 10.45  \\
sm5sf   & 50 107  & 0.137  & 0.034  & 10.72  \\
sm5qu   & 105 135 & 0.119  & 0.034  & 10.74  \\
sm6sf   & 11 151  & 0.157  & 0.033  & 10.98  \\
sm6qu   & 61 897  & 0.147  & 0.038  & 11.01  \\
sm7qu   & 8 219   & 0.157  & 0.034  & 11.29  \\
\hline
\end{tabular}
\end{center}
\end{table}

\begin{figure}
\centering
\includegraphics[width=7cm,height=7cm]{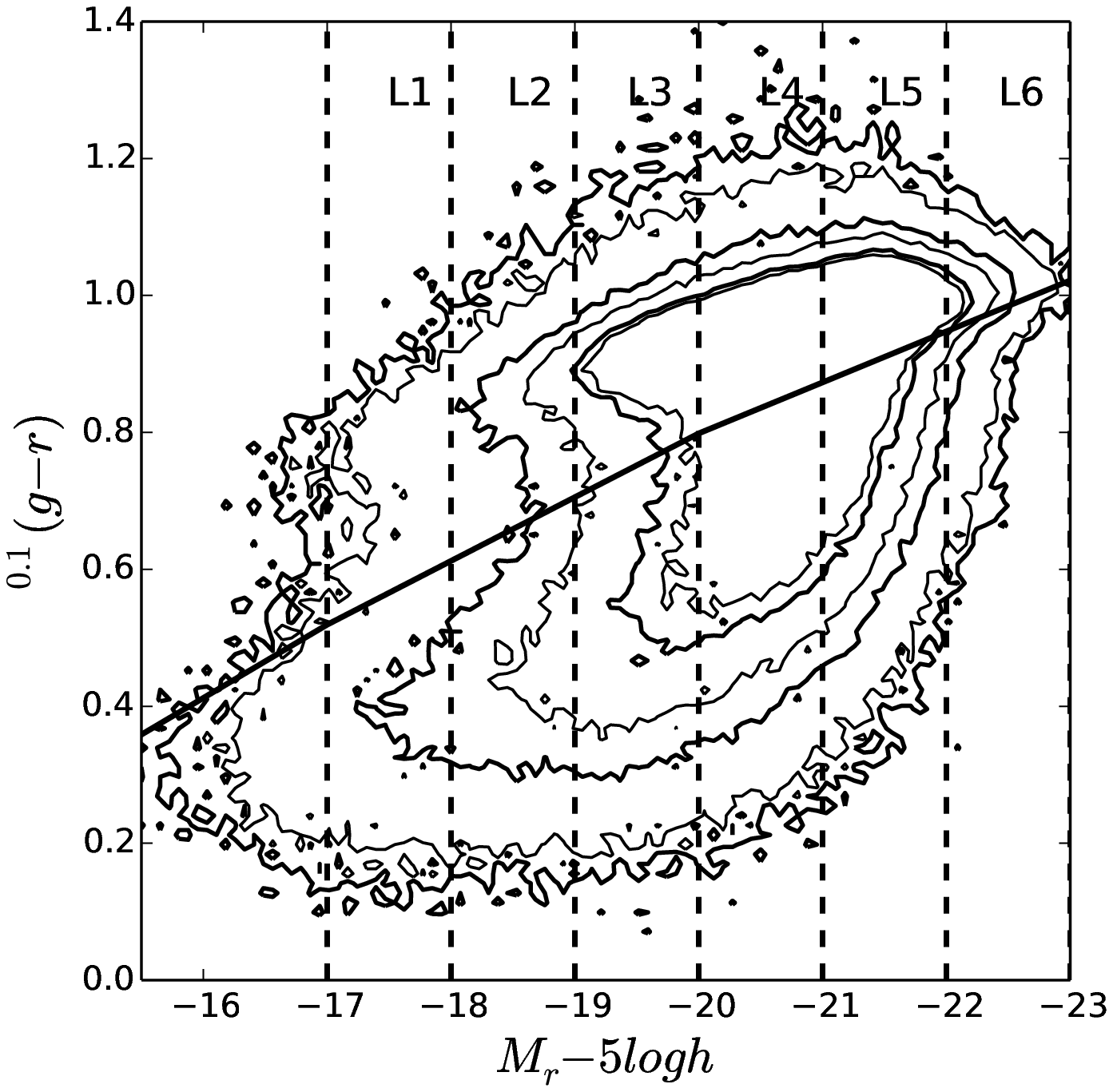}
\includegraphics[width=7cm,height=7cm]{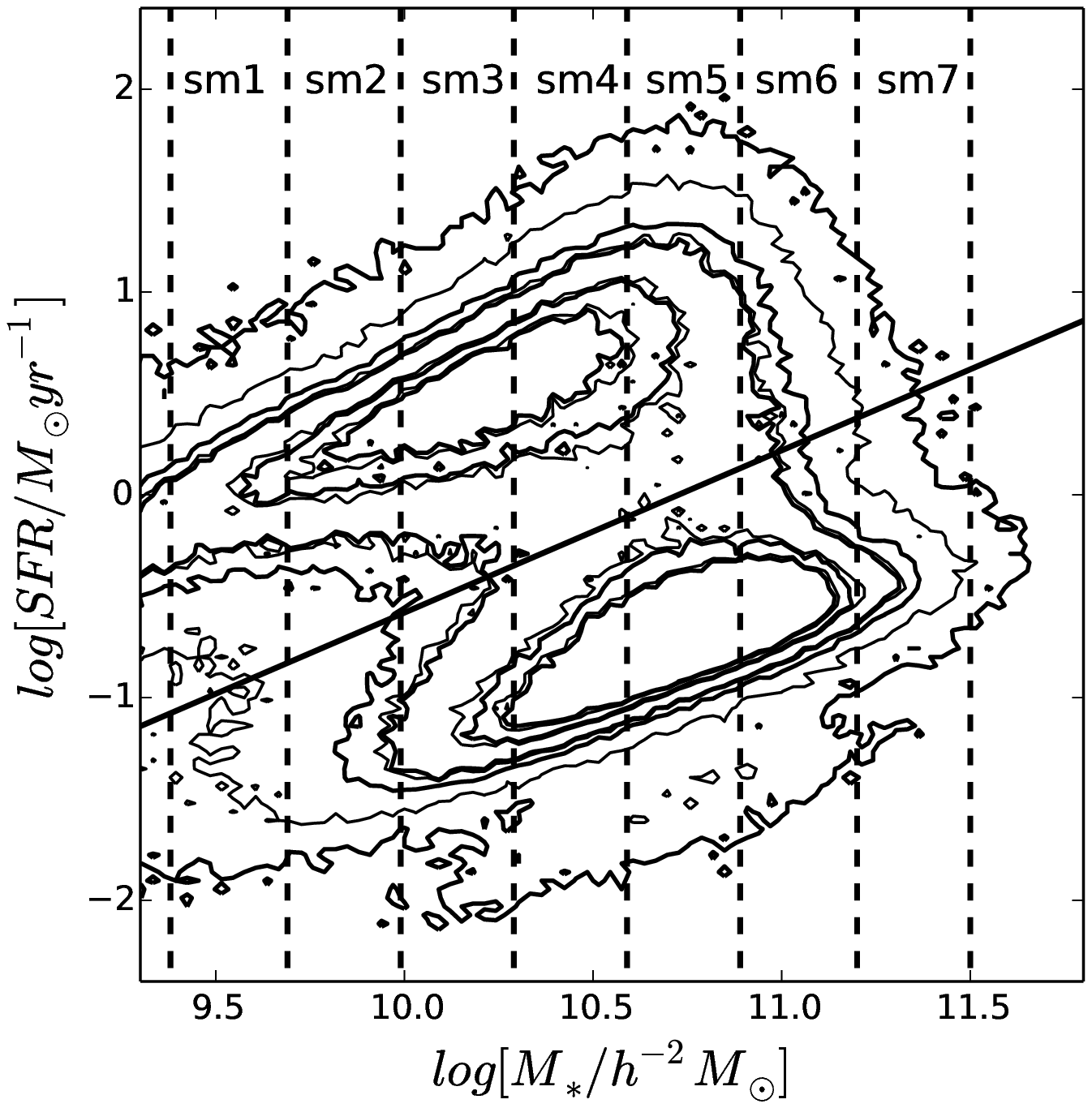}
\caption{Upper panel: The distribution of lens galaxies in the
  color-absolute magnitude plane represented by contours.  The
  luminosity bins used in the paper are shown as vertical dashed
  lines.  The solid line is the division between red and blue galaxies
  adopted from \citet{Yang2008}.  Lower Panel: The distribution of
  lens galaxies in the star formation rate (SFR)-stellar mass plane as
  represented by contours.  The stellar mass bins used in the paper
  are shown as vertical lines and the solid line is the separation
  between star forming and quenched galaxies adopted from
  \citet{Luo2014}.  }
\label{fig:sublens}
\end{figure}

\subsection{Source galaxies}

Following \citet[hereafter M05]{Mandelbaum2005}, we have only chosen
galaxies defined as OBJC\_TYPE=3 from PHOTO pipe developed by
\citet{Lupton2001}. They have to be detected both in $r$ and $i$ bands
(with $r<22$ and $i<21.6$ in model magnitudes). We first created a
preliminary catalog (further referred as Cat I) from SDSS casjobs with
115,052,555 galaxies containing positions (including {\tt run}, {\tt
  rerun}, {\tt camcol}, {\tt field}, {\tt obj}, {\tt ra}, {\tt dec}),
and photometric redshifts.

Cat I was then processed to include; (i) the sky level in unit of
photon-electron using the information of gain value in $r$ band, (ii)
the position of each galaxy in terms of CCD coordinates (in order to
get the PSF from psField files), (iii) the SPA value denoting the
angle between the camera column position with respect to north from
fpC files. We refer to this catalog as Cat II, it contains 91,941,657
galaxies. A total of 23,110,898 objects have been discarded from Cat I
either because they contain no assigned (value -9999) zero-point
extinction coefficient, airmass or sky in $r$ band, or because they
are not Flagged as BINNED1 (detected at $\geq 5$), SATURATED=0 (do not
have saturated pixels), EDGE=0 (do not locate at the edge of the CCD),
MAYBE-CR=0 (not cosmic rays), MAYBE-EGHOST=0 (not electronic ghost
line) and PEAKCENTER=0 (centroiding algorithm works well for this
object).

This pipeline was then used to process the images from fpAtlas and
psField files in order to generate our final catalog Cat III. Cat III
contains the positions, redshift, ellipticity, resolution factor and
calibration errors of each galaxy. The errors have been estimated from
both sky background and photon noise as described by Eq. 11 and Eq. 12
in M05. Only objects with valid $e_1$, $e_2$ resolution factor were
kept. As mentioned above, our pipeline will discard galaxy images with
in-convergent values of ellipticity.  From GREAT3 testing, about 40\%
galaxies were excluded due to this effect, and we further require that
${\cal R}>1/3$ which eliminates another 10\%-30\% (depending on
different simulation sets). Cat III has a final number of galaxies of
41,631,361, which is $\sim45\%$ of the original Cat II. The
Irregularity image from SDSS photo-pipe ($\sim$4\%), resolution cut
($\sim$11\%), and non-convergence ($\sim$40\%) together reduce the
number of the final catalog by $\sim$55\%.

\subsection{Lens galaxies}

We now focus on the lens galaxy sample used for this study.  Only
galaxies spectroscopically observed in the SDSS DR7 region
\citep{Abazajian2009} have been used here. More specifically, we use
the New York University Value-Added Galaxy catalog
\citep[NYU-VAGC]{Blanton2005} constructed from SDSS DR7.  All galaxies
have been extinction-corrected, with magnitudes brighter than
$r=17.72$, redshifts within the $0.01\leq z\leq 0.2$ and with a
spectroscopic redshift completeness $C_z>0.7$.  The completeness $C_z$
is defined as the average percentage of the galaxies that have
spectroscopic redshift in their local sky coverage. The resulting
galaxy sample contains a total of 639,359 galaxies for a sky coverage
of 7,748 square degrees.

In modern galaxy formation paradigm, brighter/more massive galaxies
are believed to reside in higher mass halos. This suggests that the
galaxy-galaxy lensing signals should vary with the lens galaxy
luminosity or stellar mass. Thus, a sample of brighter or more massive
lens galaxies should give a higher lensing signal. This expectation
has been proved to be correct in M05, \citet[herefater
M06]{Mandelbaum2006} and \citet{Sheldon2009}.  In M05, lens galaxies
in the SDSS DR4 are divided into six luminosity samples. We have used
the same luminosity binning for our SDSS DR7 galaxies.  The selection
criteria and galaxy numbers of our six lens galaxy samples are listed
in Table~\ref{tab:tbl-1}. The scatter of the redshift distribution,
the ratio between the mean luminosity and the characteristic
luminosity $L_*$ ($M_*=-20.44$, as given in \cite{Blanton2003}), and
the number of galaxies contained in each sample are also listed in
Table. \ref{tab:tbl-1}.  On average, the number of galaxies in our
sample is 2 to 3 times larger than the corresponding M05 sample,
simply because DR7 covers a larger area than DR4 (7748 v.s.  4783
square degrees). The mean redshift from our lens sample is slightly
lower than that of M05, because M05 also used lenses at $z>0.2$ while
the redshift range of our sample is between $0.01$ and $0.2$. The
redshift distributions of our lens samples are shown in the upper
panel of Fig.~\ref{fig:zlens}. The solid black line is for the total
sample, while the colored lines are for the six luminosity samples.

We further divide galaxies in each luminosity bin into blue and red
sub-samples according to
\begin{equation}
^{0.1}(g-r)=1.022-0.0652x-0.0031x^2 \,,
\label{eq:eq_color}
\end{equation}
where $x=\rmag + 23.0$ \citep{Yang2008}. The upper panel of 
Fig.\,\ref{fig:sublens} shows the distribution of the lens galaxies 
in the color-absolute magnitude plane, with the black dashed line 
showing the demarcation line (Eq. \ref{eq:eq_color}), and
the vertical lines marking the different luminosity bins we use.

In M06, galaxy-galaxy lensing signals are measured for lens galaxies
binned in stellar masses. Here we make a similar binning for our SDSS
DR7 galaxies.  Note, however, that the stellar masses in M06 are
estimated from galaxy spectra, as described in \citet{Kauffmann2003},
while the stellar masses in our sample are estimated using the model
described in \citet{Bell2003}. Table~\ref{tab:tb2-1} lists the general
properties of our samples in different stellar mass bins, such as the
number of galaxies in our samples in comparison to that in M06, the
mean redshift, the scatter in redshift, and the mean stellar mass.
Shown in the lower panel of Fig. \ref{fig:zlens} are the redshift
distributions of our lens samples in different stellar mass bins.  The
solid black line is for the total sample, while the colored lines are
for the seven stellar mass samples, as indicated.

We further divide galaxies in each stellar mass bin into red and blue
sub-samples using Eq.\,(\ref{eq:eq_color}).  Table~\ref{tab:tb2-2}
shows the number, mean redshift, scatter in redshift, and the mean
stellar mass of the galaxies in each of the color sub-samples. In
general, the mean stellar mass of the red sample is larger than that for
 the corresponding blue sample by 0.01 to 0.02 dex.

In addition to the color separation, we also separate galaxies in
different stellar mass bins into star-forming and quenched
sub-samples. Here we use the scheme given in  
\citet{Yang2013, Luo2014} to define the star-forming and quenched 
populations, and the dividing line is defined to be
\begin{equation}
\log SFR=(\log M_* -2\log h -11.0)\times 0.8\,.
\label{eq:eq_sf_qu}
\end{equation}
The lower panel of the Fig. \ref{fig:sublens} shows the distribution
of galaxies in the SFR - stellar mass plane, with the black line
showing the division defined in Eq. \ref{eq:eq_sf_qu}. Note that $M_*$
is presented in units of $\msunhh$.  Table~\ref{tab:tb2-3} lists the
number, the mean redshift, the scatter in redshift, and the mean
stellar mass of each subsample.  For each mass bin, the average
stellar masses in the two subsamples are similar, while the mean
redshifts differ slightly, with the quenched subsample has a slightly
higher mean redshift than the corresponding star-forming subsample.

\subsection{Galaxy-galaxy lensing signals}

From weak lensing shear measurements, we can estimate the excess
surface density (ESD) of the lens system, which is defined as
\begin{equation}
\Delta\Sigma(R)=\Sigma(\leqslant R)-\Sigma(R)\,.
\end{equation}
Here $\Sigma(\leqslant R)$ and $\Sigma(R)$ are the mean surface mass
density inside a certain radius $R$ and at the radius $R$,
respectively. The tangential shear is related to this quantity via a
critical density,
\begin{equation}
\gamma_t(R)\Sigma_c=\Delta\Sigma(R)\,,
\label{eq:gamma_R}
\end{equation}
where the  critical density in a lensing system is,
\begin{equation}
\Sigma_c^{-1}=\frac{4\pi G}{c^2}\frac{D_lD_{ls}(1+z_l)^2}{D_s}\,,
\end{equation}
with $D_s$, $D_l$ and $D_{ls}$ being the angular diameter distances of
the source, the lens and between the lens and the source,
respectively.

The mean excess surface density around a lens galaxy is related to the
line-of-sight projection of the galaxy-matter cross correlation
function,
\begin{equation}
\xi_{\rm gm}(r)=\langle\delta({\bf x})_{g}\delta({\bf x}+{\bf r})_{m}\rangle, 
\end{equation}
so that
\begin{equation}
\label{sigatr}
\Sigma(R) = 2 \overline{\rho} \int_{R}^{\infty} \xi_{\rm gm}(r) 
{r \, \rmd r \over \sqrt{r^2 - R^2}}\,,
\end{equation}
and
\begin{equation}
\label{siginr}
\Sigma(\leq R) = \frac{4\overline{\rho}}{R^2} \int_0^R y\,\,dy\,
 \int_{y}^{\infty} \xi_{\rm gm}(r) {r \, \rmd r \over \sqrt{r^2 - y^2}}\,
\end{equation}
where $\overline{\rho}$ is the average background density of the
Universe.  Note that in both equations, we have omitted the
contribution from the mean density of the universe, as it does not
contribute to the ESD.

In order to take into account source galaxy photometric redshift errors,
it is necessary to convolve the results with the error distribution (see M05),
\begin{equation}
\Sigma_c^{-1}(z_l,z_p)=\int p(z_s|z_p)\Sigma_c^{-1}(z_l,z_s)dz_s\,,
\end{equation}
where $z_l$, $z_p$, $z_s$ are the spectroscopic redshift of the lens
galaxy, the photometric redshift of the source galaxy and the
spectroscopic redshift of the source galaxy, respectively.  Since the
spectroscopic redshifts are not available for most source galaxies,
the determination of $p(z_s|z_p)$ relies on other spectroscopic
surveys. We follow M05 and use the error distribution obtained by
cross identifying the subsample of their source galaxies with other
spectroscopic surveys such as DEEP2, COMBO-17.

%%% Luminosity Bins All-----------------------------------------------
\begin{figure}
\centering
\includegraphics[width=7cm,height=9cm]{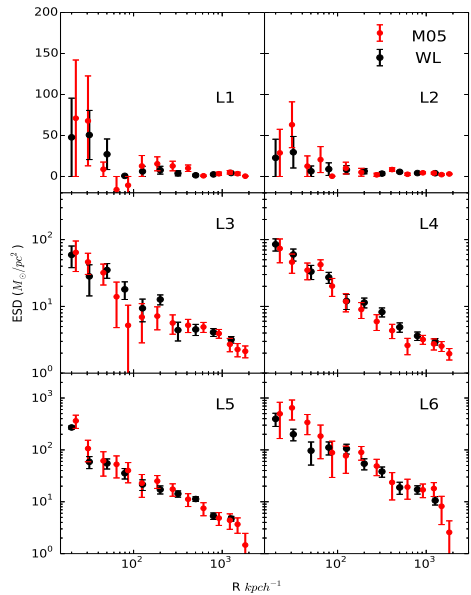}
\caption{Excess surface density (ESD) of our lens galaxies in six
  luminosity bins. The black dots are our measurements and the red
  dots are results obtained by M05.}
  \label{fig:compare_L}
\end{figure}

\begin{figure}
\centering
\includegraphics[width=7cm,height=9cm]{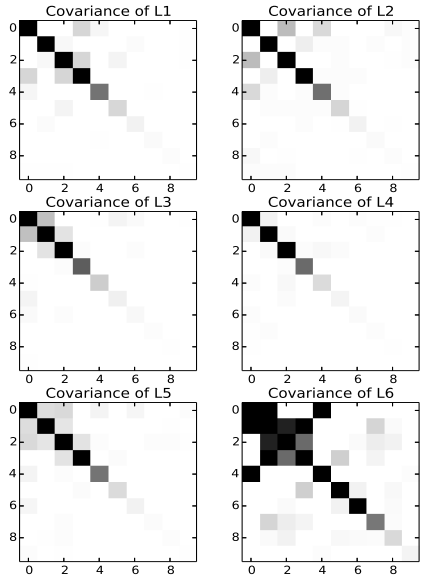}
\caption{Covariance matrix of the data points for our six
  luminosity bins. The grey scale color has been scaled so that
  smaller values are reflected on this covariance map.  The values
  of the covariance map are provided in separate files. }
  \label{fig:covar_L}
\end{figure}

Fig. \ref{fig:compare_L} shows the average excess surface density of
our lens galaxies divided into six luminosity bins. The black dots are
our measurements and the red dots shows the M05 data (kindly provided
by Rachel Mandelbaum). For simplicity, the signals around each galaxy
sample were calculated in 10 equal logarithmic bins rather than 45 bin
then re-binned as in M05. The error bars are estimated using 2500
bootstrap resampling of the lens galaxy samples. The covariance matrix
of the data points shown in Fig.  \ref{fig:compare_L} are given in
Fig.  \ref{fig:covar_L}.  We rescaled the color so that smaller values
can be seen. Interested readers can find the covariance values via the
link provided at the end of Section \ref{sec_intro}. The ESDs for lens
galaxies in different luminosity bins are also listed in Table
\ref{tab:ESD-L6} in Appendix \ref{sec_ESD}.

As in M05, we provide a detailed list of possible systematic errors in
the measurements in Appendix \ref{sec_systematic}.  The total possible
$2\sigma$ systematic error in terms of $\delta \gamma/\gamma$ is about
$[-9.1\%,20.8\%]$. This is roughly consistent with those quoted in
M05, about $[-9.0\%,+18.4\%]$, as we are using roughly the same
selection criteria for source galaxies.  In addition, the redshift
tests (using foreground galaxies as sources) and $\gamma_{45}$
component tests are consistent with zero.  Note that these possible
systematic errors are mainly associated with the type of source
galaxies that are used, where a brighter magnitude cut will reduce the
systematics significantly.  On the other hand, the total number of
galaxies that are used in our investigation impacts the statistical
errors. As one can see in Fig. \ref{fig:compare_L}, our results are in
good agreement with M05, however with much smaller error bars, since
we have larger number of lens galaxies in our SDSS DR7 galaxy
samples. There is a clear trend that the amplitude of $\Delta\Sigma$
increases as the luminosity increases.

%%% Luminosity Bins Red and Blue-------------------------------------------
\begin{figure}
\centering
\includegraphics[width=7cm,height=9cm]{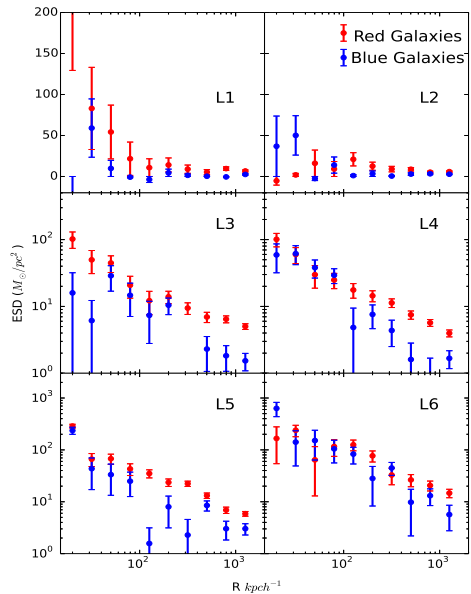}
\caption{The ESDs for  red (red dots) and blue (blue dots) galaxies in
  different luminosity bins.}
  \label{fig:Lbin_color}
\end{figure}

For each of our luminosity bin, we also obtain the galaxy-galaxy
lensing signals separately for the red and blue subsamples, and the
results are presented in Fig. ~\ref{fig:Lbin_color}.  The error bars
here are larger due to the decreased number of lens galaxies per
subsample. For very faint lens galaxies in the L1 bin, the red
galaxies have larger ESDs than blue galaxies especially at small
radius. This indicates that faint red galaxies tend to be located in
relatively more massive halos than their blue counterparts.  For
brighter galaxies, especially in L2-L4 bins, the red and blue galaxies
show similar ESDs at small scales (with the caveat that the error bars
are big), but red galaxies have much higher amplitudes than their blue
counterparts at $R> 200 \kpch$. The latter indicates that these red
galaxies are preferentially located in high density regions.

%%% Stellar Mass Bins Red and Blue-----------------------------------------------

\begin{figure}
\centering
\includegraphics[width=7cm,height=9cm]{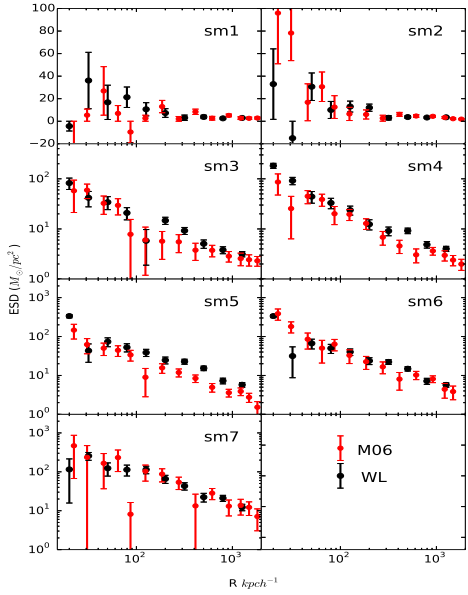}
\caption{ESDs for lens galaxies of different
  stellar masses. In each panel we compare our results 
  (black dots) with those of M06 (red dots). }
  \label{fig:compare_sm}
\end{figure}

We have also estimated the ESDs for galaxies in different stellar mass
bins, and the results are shown in Fig.~\ref{fig:compare_sm} with
black dots, in comparison with the results of M05 that are shown as
the red dots. Here again our results agree with those of M06, except
in the sm5 bin where our results at $200 < R < 1,500 \kpch$ are
significantly (by a factor of about two) higher. Since our sample is
larger than that of M06 (SDSS DR7 v.s. DR4), this enhancement
indicates that a significant portion of the additional galaxies in our
sample may be located in or near massive structures.  Once again, we
provide the ESDs for lens galaxies in different stellar mass bins in
Table \ref{tab:ESD-sm7} in the Appendix \ref{sec_ESD}.

%%% Stellar Mass Bins SF and QU------------------------------------------------------

\begin{figure}
\centering
\includegraphics[width=7cm,height=9cm]{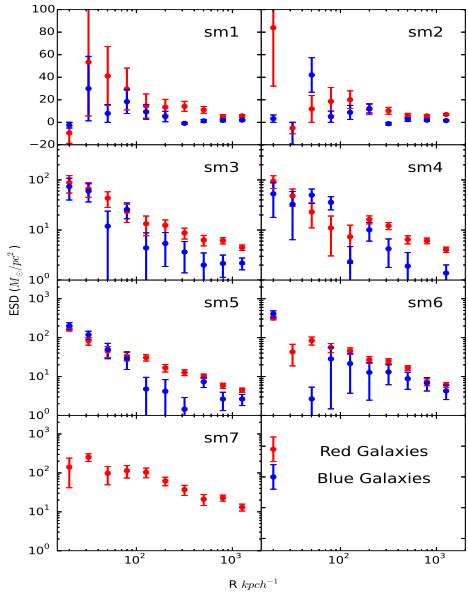}
\caption{ESDs for red (red dots) and blue (blue dots) lens
  galaxies in  different stellar mass bins.}
  \label{fig:sm_color}
\end{figure}

\begin{figure}
\centering
\includegraphics[width=7cm,height=9cm]{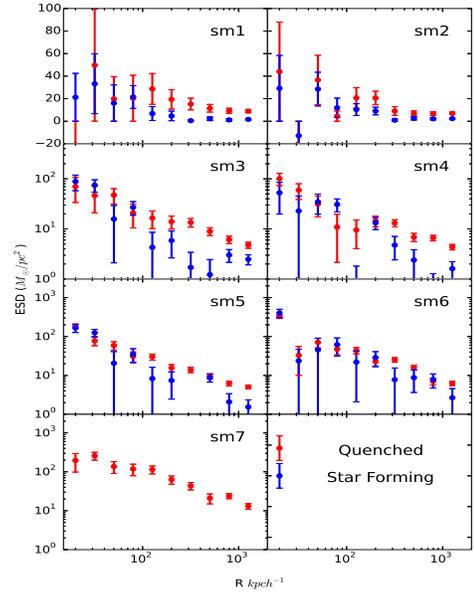}
\caption{ESDs for quenched (red dots) and star forming (blue dots)
  lens galaxies in different stellar mass bins.}
  \label{fig:sm_sq}
\end{figure}

Finally, we also measure the ESDs for our star-formation subsamples of
color and of star formation in stellar mass bins.
Fig. \ref{fig:sm_color} shows the results for red versus blue
galaxies. The color dependencies in different stellar mass bins are
quite similar to those in different luminosity bins. In addition, as
the color of a galaxy may be related to the star formation history of
the galaxy, the dependence on star formation shown in
Fig.\ref{fig:sm_sq} is similar to the color dependence.

The galaxy-galaxy lensing signals can be fitted to obtain the average
halo mass of the lens systems. With the results we obtained here, we
will be able to study how galaxies of different properties
(luminosity, stellar mass, color and star formation) are linked to
dark matter halos.  However, as pointed out in \citet{Yang2006a} and
found in \citet{Li2014} and \citet{Li2016}, the central and satellite
galaxies have very different lensing signals.  It is thus important to
separate samples into centrals and satellites in order to model the
observed ESDs in detail.  We will come back to this in a forthcoming
paper.

\section{Summary and Discussion}
\label{sec_summary}

In weak lensing studies, obtaining a reliable measurement of the
lensing signals requires highly accurate image processing. In this
paper, we build our image processing pipeline to achieve accurate
shape measurement for weak lensing studies based on \citet[
(BJ02)]{Bernstein2002} and \citet[(HS03)]{Hirata2003} methods.  This
pipeline is then applied to SDSS DR7 to measure the galaxy shapes, as
well as the galaxy-galaxy lensing signals for lens galaxies of
different luminosities, stellar masses, colors, and SFRs. The main
results of this paper are summarized as follows.
\begin{itemize}
\item We have developed a new image processing pipeline, and tested it
  on SHERA and GREAT3 simulations.  Our pipeline works well on PSF
  correction in the absence of sky background noise. The corrected PSF
  multiplicative errors are far below the 1\% requirements ($0.009\%$
  for $\gamma_1$ and $-0.053\%$ for $\gamma_2$) for PSF correction
  only.

\item An non-convergence problem occurs for $\sim 40\%$ galaxies when
  more realistic simulations with sky background noise are being used.
  In addition, to have a sufficient image resolution ${\cal R}>1/3$ an
  additional 20\% have to be discarded. Despite these, our method
  achieves a lensing reconstruction accuracy that is similar to other
  methods as shown in the GREAT3 competition \citep{Mandelbaum2015}.

\item Our pipeline was applied to the SDSS DR7 $r$ band imaging data
  and create a catalog containing 41,631,361 galaxies with information
  about position, photometric redshift, ellipticity and ellipticity
  measurement error due to sky background and Poisson noise.

\item Using these galaxy images, we calculated the galaxy-galaxy
  lensing signals around foreground lens galaxies binned in different
  luminosities and stellar masses. Our results show good agreement
  with the previous studies of \citet[M05]{Mandelbaum2005} and
  \citet[M06]{Mandelbaum2006}, with significantly reduced error bars.

\item We have also separate the galaxies in different luminosity/
  stellar mass bins into red/blue or star-forming/quenched
  subsamples. The galaxy-galaxy lensing signals show quite different
  scale dependences among these subsamples. While red and quenched
  galaxies show stronger galaxy-galaxy lensing signals than their
  counterparts in the same luminosity or stellar mass bins, the
  enhancement is the strongest at relatively large separations.
\end{itemize}

As the first paper of our galaxy-galaxy lensing series, here we have
focussed on testing the reliability of our image processing pipeline
and presented some general results of the galaxy-galaxy lensing in the
SDSS DR7. In addition, we have performed a number of tests on possible
systematics in our pipeline, using the $\gamma_{45}$ component,
foreground sources, and random samples. Our pipeline and the
galaxy-galaxy lensing signals obtained prove to be reliable against
these tests.

Our data can be used to study the dark matter contents associated with
SDSS galaxies and the structures they represent.  In a forthcoming
paper, we will use the data to carry out a number of analyses.  We
will separate galaxies into centrals and satellites so as to model the
mass distributions around them and their links to dark matter halos.
We will also obtain the mass distribution around galaxy groups
\citep{Yang2007} to test the reliability of the mass assignments based
on other mass estimates, and to study how halo masses depend on the
intrinsic properties of galaxy groups, such as the colors of members
of galaxy groups. Finally we will stack the lensing signals around
groups with different X-ray properties \citep[e.g.][]{Wang2014}
to test how X-ray gas in galaxy groups is related to their dark matter
contents.

As we have found, our pipeline is unable to fully deal with images
that are noisy. This limitation is the main drawback of our pipeline
and needs to be addressed. Fourier space based methods seem to be
superior in this regard as they can process asymmetric systems and
much noisier images. For this reason, we intend to improve our
methodology by implementing the Fourier space method of
\citet{Zhang2015}.

\acknowledgements

WL thanks Rachel Mandelbaum from Carnegie Mellon University for very
useful guiding and discussions at various stages of this project and
providing the data points presented in this paper.  WL also thanks
Dandan Xu from Heidelberg University for useful discussion.  This work
was supported by the following programs; the 973 Program
(No. 2015CB857002), NSFC (Nos. 11128306, 11121062, 11233005,
11503064), the Strategic Priority Research Program ``The Emergence of
Cosmological Structures" of the Chinese Academy of Sciences, Grant No.
XDB09000000, and a key laboratory grant from the Office of Science and
Technology, Shanghai Municipal Government (No. 11DZ2260700) as well as
Chinese Scholarship Council (201504910477) and Shanghai Natural
Science Foundation, Grant No. 15ZR1446700. LF acknowledges support
from NSFC grants 11103012 \&11333001 \& Shanghai Research grant
13JC1404400. LR acknowledges the NSFC(grant No.11303033), the support
from Youth Innovation Promotion Association of CAS.

This work was also supported by the High Performance Computing
Resource in the Core Facility for Advanced Research Computing at
Shanghai Astronomical Observatory.

%%%%%%%%%%%%%%%%%
% Appendix
%%%%%%%%%%%%%%%%%

\appendix

\section{ BJ02 Method}
\label{bj02_detail}

We use this first appendix to detail the mathematical derivation of
the PSF anisotropy correction of the pipeline presented in this paper.
We follow BJ02 using  the following eigenfunction expansion for our
images,
\begin{equation}\label{eq:kernel}
K =\sum_{kl}k_{kl}D_{kl}, 
\end{equation}
where 
\begin{eqnarray} \label{eq:Dkl}
D_{kl}&=&\left(\frac{\partial}{\partial x}+i\frac{\partial}{\partial y}\right)^k
       \left(\frac{\partial}{\partial x}-i\frac{\partial}{\partial y}\right)^l \nonumber\\
       &=&\sigma^{-(k+l)}(a_q^{\sigma\downarrow}-a_p^{\sigma\uparrow})^k(a_p^{\sigma\downarrow}-a_q^{\sigma\uparrow})^l.
\end{eqnarray}
Note that $k,l$ in $D_{kl}$ are the index of matrix components, while on the right hand sides
they represent power indices.
The operators $a_{p}^{\downarrow}$ and $a_{p}^{\uparrow}$ are the 
lowering and raising operators for the 2D QHO eigenfunctions, which have
the properties that
\begin{eqnarray}
a_{p}^{\downarrow}&=&\frac{1}{2}\left[\frac{x-iy}{\sigma}+\sigma\left(\frac{\partial}{\partial x}-i\frac{\partial}{\partial y}\right)\right]\,,\nonumber\\
a_{p}^{\uparrow}&=&\frac{1}{2}\left[\frac{x+iy}{\sigma}-\sigma\left(\frac{\partial}{\partial x}+i\frac{\partial}{\partial y}\right)\right]\,,\nonumber\\
a_{q}^{\downarrow}&=&\frac{1}{2}\left[\frac{x+iy}{\sigma}+\sigma\left(\frac{\partial}{\partial x}+i\frac{\partial}{\partial y}\right)\right]\,,\nonumber\\
a_{q}^{\uparrow}&=&\frac{1}{2}\left[\frac{x-iy}{\sigma}-\sigma\left(\frac{\partial}{\partial x}-i\frac{\partial}{\partial y}\right)\right]\,.
\end{eqnarray}
The matrix $D_{kl}$ defined above makes it easy to use the raising and lowering
operators to determine how a given kernel will act on an image.  As we
are dealing with discrete image data, the derivative along $x$ and
$y$ axis can be treated as convolving a $3\times 3$ matrix with the
image $I(x,y)$,
\begin{eqnarray}
\frac{\partial I}{\partial x}&=&\begin{pmatrix}
0 & 0 &0\\ 
\frac{-1}{2} & 0 & \frac{1}{2} \\
0 & 0 & 0
\end{pmatrix}\otimes I\,,\nonumber\\
\frac{\partial I}{\partial y}&=&\begin{pmatrix}
0 & \frac{1}{2} &0\\
0  & 0 & 0\\
0 & \frac{-1}{2} & 0
\end{pmatrix}\otimes I\,,\nonumber\\
\frac{\partial^2 I}{\partial x^2}&=&\begin{pmatrix}
0 & 0 & 0\\
1  & -2 & 1\\
0 & 0 & 0
\end{pmatrix}\otimes I\,,\nonumber\\
\frac{\partial^2 I}{\partial y^2}&=&\begin{pmatrix}
0 & 1 &0\\
0  & -2 & 0\\
0 & 1 & 0
\end{pmatrix}\otimes I\,,\nonumber\\
\frac{\partial^2 I}{\partial x \partial y}&=&\begin{pmatrix}
\frac{-1}{4} & 0 &\frac{1}{4}\\ 
0  & 0 & 0\\ 
\frac{1}{4} & 0 & \frac{-1}{4}
\end{pmatrix}\otimes I\,.
\end{eqnarray}
These are all the components up to the Second Derivative in the
Gradient Direction (SDGD) one can get  from the $3\times 3$ discrete
image pixels.   The related $D_{kl}$ in Eq. \ref{eq:Dkl} are:
\begin{align}
D_{10}=\begin{pmatrix}0 & i(1/2) &
0\\ -1/2 & 0 & 1/2\\0 & i(-1/2) & 0\end{pmatrix}; D_{01}=\overline{D_{10}},\\
D_{20}=\begin{pmatrix}i(-1/2) & -1 &
i(1/2)\\ 1 & 0 & 1\\i(1/2) & -1 & i(-1/2)\end{pmatrix}; D_{02}=\overline{D_{20}},\\
D_{11}=\begin{pmatrix}0 & 1 &
0\\ 1 & -4 & 1\\0 & 1 & 0\end{pmatrix}.
\end{align}
\begin{eqnarray}
D_{10}&=&\begin{pmatrix}
0 & i(1/2) &0\\ 
-1/2 & 0 & 1/2\\
0 & i(-1/2) & 0
\end{pmatrix}=\overline{D_{01}}\,,\nonumber\\
D_{20}&=&\begin{pmatrix}
i(-1/2) & -1 &i(1/2)\\ 
1 & 0 & 1\\
i(1/2) & -1 & i(-1/2)
\end{pmatrix}=\overline{D_{02}}\,,\nonumber\\
D_{11}&=&\begin{pmatrix}
0 & 1 &0\\
1 & -4 & 1\\
0 & 1 & 0
\end{pmatrix}\,.
\end{eqnarray}
Note that $D_{00}$ is the identical matrix, and $D_{11}$ is actually a
Laplacian operator.  The components listed above contain all the first
and second order derivatives.  Higher order derivatives can be
obtained by convolving the above $3\times 3$ components. For instance,
$D_{22} = D_{20}\otimes D_{02}$.  Note that since $D_{ij}$ are
complex, to end up as a real image, $k_{ij}$ are required to satisfy
$k_{ij}=\overline{k_{ij}}$.

Combining equations \ref{eq4} and \ref{eq:kernel} we have
\begin{equation}
\mathrm{b}^{*}=\sum_{ij}k_{ij}D_{ij}\mathrm{b},
\end{equation}
where $D_{ij}\mathrm{b}$ obeys the recursion:
\begin{eqnarray}
\label{eq:recursive}
D_{00}\mathrm{b}&=&\mathrm{b}\,,\nonumber\\
D_{(i+1)j}\mathrm{b}&=&\frac{1}{\sigma}(a_{q}^{\downarrow}-a_{p}^{\uparrow})D_{ij}\mathrm{b}\,,\nonumber\\
D_{i(j+1)}\mathrm{b}&=&\frac{1}{\sigma}(a_{p}^{\downarrow}-a_{q}^{\uparrow})D_{ij}\mathrm{b}\,.
\end{eqnarray}

The final step is to constrain the coefficients $k_{ij}$ by requiring 
$\mathrm{b}^{*}$ to meet the requirements,
\begin{equation}
\mathrm{b}_{pq}^{*}=0, (m=p-q=2).
\end{equation}
We construct a $5\times 5$ kernel to remove the anisotropy. For
simplicity, however, we demonstrate the procedure by reconstructing a
$3\times 3$ kernel to an upper limit $p+q\le N=4$. Increasing to
higher order expansion does not improve our results significantly.
The coefficient matrix used to constrain $k_{ij}$ is then
\begin{equation}
\mathrm{b}^{*}=\begin{pmatrix} \mathrm{b}^{*}_{00}  & \mathrm{b}^{*}_{01}  &
\mathrm{b}^{*}_{02} \\ \mathrm{b}^{*}_{10}  & \mathrm{b}^{*}_{11}  & \mathrm{b}^{*}_{12} \\
\mathrm{b}^{*}_{20} & \mathrm{b}^{*}_{21}  & \mathrm{b}^{*}_{22} \end{pmatrix}.
\end{equation}
In the ideal case, we have
\begin{equation}
\mathrm{b}^{*}=\begin{pmatrix} \frac{1}{\sqrt{\pi}}  & 0  &
0 \\ 0  & -\frac{1}{\sqrt{\pi}}  & 0 \\
0 & 0  & \frac{1}{\sqrt{\pi}} \end{pmatrix}.
\end{equation}
Since $\mathrm{b}^{*}_{10}$ naturally goes to zero if the PSF's
centroid is measured accurately, this term does not have any
constraining power on $k_{ij}$. For the unspecified
$\mathrm{b}_{pq}^*$, e.g. $\mathrm{b}_{31}^*$, $\mathrm{b}_{40}^*$, we
set $k_{pq}=0$ as in BJ02 while still meet the kernel requirement to
remove the PSF anisotropy to some order. The components that remain
are only $\mathrm{b}^{*}_{00}$, $\mathrm{b}^{*}_{11}$,
$\mathrm{b}^{*}_{20}$ and $\mathrm{b}^{*}_{22}$. The linear equation
to calculate $k_{ij}$ is then
\begin{equation}
\begin{pmatrix} D_{00}b_{00}  & D_{01}b_{00} & D_{02}b_{00} & D_{11}b_{00} \\ 
                D_{00}b_{11}  & D_{01}b_{11} & D_{02}b_{11} & D_{11}b_{11} \\
                D_{00}b_{11}  & D_{01}b_{20} & D_{02}b_{20} & D_{11}b_{20} \\
                D_{00}b_{22}  & D_{01}b_{22} & D_{02}b_{22} & D_{11}b_{22}\end{pmatrix} 
                \begin{pmatrix} k_{00} \\ k_{10} \\
                 k_{02} \\ k_{11} \end{pmatrix} =\begin{pmatrix} b^*_{00} \\ b^*_{11} \\
                 b^*_{20} \\ b^*_{22} \end{pmatrix}=
                 \begin{pmatrix} \frac{1}{\sqrt{\pi}}  \\ -\frac{1}{\sqrt{\pi}}  \\
                 0 \\ \frac{1}{\sqrt{\pi}}  \end{pmatrix}
\end{equation}
Owing to the fact that $k_{kl}=\bar{k_{kl}}$, the dimensions shrink
dramatically while considering $k_{01}$ and $k_{02}$.  Before solving
this linear equation, we have to calculate each elements of the
coefficient matrix. $D_{kl}\textbf{b}$ denotes all the entries of the
coefficient vector when expanding the PSF image using elliptical
Laguerre polynomials. In practice, for the pixellized image data,
$D_{kl}\textbf{b}$ can be written as follows,
\begin{equation}
D_{10}\textbf{b}=\frac{1}{2}(T_{z1}\textbf{b}-T_{-z1}\textbf{b})+
\frac{1}{2}i(T_{z2}\textbf{b}-T_{-z2}\textbf{b})
\end{equation}
where $z1=1/\sigma$, $z2=i/\sigma$. $T_z$ is defined as the translation operator.
% \begin{gather}
%   T_zf(x,y)=f(x-x_0,y-y_0),\\
%   z=(x_0+iy_0)/\sigma.
% \end{gather}
\begin{eqnarray}
  T_zf(x,y)&=&f(x-x_0,y-y_0) \,, \nonumber\\
  z&=&(x_0+iy_0)/\sigma\,.
\end{eqnarray}
So we have
% \begin{gather}
% T_{z1}f(x,y)=f(x-1,y),\\
% T_{z2}f(x,y)=f(x,y-1).
% \end{gather}
\begin{eqnarray}
T_{z1}f(x,y)=f(x-1,y) \,, \nonumber\\
T_{z2}f(x,y)=f(x,y-1)\,.
\end{eqnarray}
The functional form of $T_z$ can be derived from the decomposition of
PSF image, i.e. $P=\sum b_{pq}\psi_{pq}^{\sigma}$ and $T_{z}P=\sum
b_{pq}^{'}\psi_{pq}^{\sigma}$. We denote $\textbf{b}^{'}$ as the new
coefficients after operation $T_z$
\begin{eqnarray} 
\textbf{b}^{'}&=&T_z\textbf{b}\,, \nonumber\\
b_{p'q'}^{'}&=&\sum T_{p'q'}^{pq}b_{pq}\,, \nonumber\\
T_z\psi_{pq}^{\sigma}&=&\sum T_{p'q'}^{pq}\psi_{p'q'}^{\sigma}\,, \nonumber\\
T_{p'q'}^{pq}&=&\sigma^{2}\int d^2x(T_{z}\psi_{pq}^{\sigma})\bar{\psi_{p'q'}^{\sigma}}\,.
\end{eqnarray}
This directly leads to the first term,
\begin{equation}
T_{00}^{00}=e^{-|z|^2/4}.
\end{equation}
The left terms can also be solved recursively with the following relation,
\begin{eqnarray}
T_{p'q'}^{pq}&=&h(p,p')\bar{h}(q,q') \,, \nonumber\\
h(p,0)&=&\frac{(-z/2)^p}{\sqrt{p!}}e^{-|z|^2/8}\,, \nonumber\\
h(p,p'+1)&=&[\sqrt{p}h(p-1,p')+\frac{1}{2}\bar{z}h(p,p')]/\sqrt{p'+1}\,.
\end{eqnarray}

\section{ Systematic checks}
\label{sec_systematic}

We present, in this appendix, the main systematic errors relevant to
our study and the test we performed to check for additional systematics. 

\subsection{Systematic errors}

There are five major systematics in weak lensing measurement as
described in HS03 and M05. Table. \ref{tab:tbl-6} lists these major
biases in our work and compare them to M05. We give below a brief
introduction for each of them.

\begin{table}[h!]
\begin{center}
  \caption{\label{tab:tbl-6} This table lists the five major
    systematics in our weak lensing measurements compared to M05. }
\begin{tabular}{ccc}
\hline
Bias (per cent) & M05 & this work\\
\hline
Selection bias            & [0, 10.3]               & [0, 12.3] \\
PSF reconstruction bias   & $\pm 2.1$ to $\pm 2.4$ & $\pm $2.2 \\
PSF dilution bias         & [-2.8, 4.0]             & *[-2.8, 4.0] \\
Shear responsivity error  & [0, 1.7]                & [0, 2.3]\\
Noise rectification error & [-3.8, 0]               & [-4.08, 0] \\
Total $2\sigma$ $\delta \gamma/\gamma$(per cent) & [-9.0,18.4] & [-9.1,20.8]\\
\hline
\end{tabular}
\end{center}
\end{table}

\subsubsection{Selection bias}

The first selection bias is mainly caused by the asymmetries of the
PSF, denoted as `PSF selection bias' in \citet{Kaiser2000}. More
galaxies are selected if they are elongated in one direction.
Secondly, the shear introduce asymmetries in the same way as the
PSF. Shear stretches galaxies along a certain direction and hence
makes the major axes of galaxies aligned with that direction more
easily detected. In HS03, this is referred to as the `shear selection
bias'. Finally, many significance-based object detection methods
preferentially select circular objects leading to underestimation of
the shear signal. M05 estimates the selection bias to be [0, 5.7]\%
for galaxies with $r<21$, [0, 10.3]\% for $r>21$ and [0, 11.1]\% for
LRG samples. Both M05 and this work directly use the catalog from
PHOTO pipeline, and the selection bias from M05 and our catalog will
not differ from each other significantly.  Following Eq.19 in M05, the
selection bias is calculated as
\begin{equation}
\frac{\delta \gamma}{\gamma}=
\frac{\bar{R}_{min}(1-\bar{R}_{min})}{\bar{R}}e_{rms}^2n(\bar{R}_{min})
\end{equation}
where $\bar{R}$ is the shear responsiveness and
$n(\bar{R}_{min})=1.6$, 2.4 and 2.8 for $r<21$, $r>21$ and LRG samples
in M05, respectively.  We have used the value for $r>21$ here,
estimating $n(\bar{R}_{min})$ for our sample to be 2.4. The maximum
possible systematics can be induced by selection bias is $12.3\%$ in
our sample, slightly larger than those obtained by M05.

\subsubsection{PSF reconstruction bias}

This bias arises from the process of reconstructing the PSF from the
PHOTO PSF pipeline. This bias estimated in M05 is $\pm 2.1$ to $\pm
2.5$ for SDSS sample. Since the PSF applied in M05 and this work are
both from PHOTO PSF pipeline, we follow M05 and also use Eq. 20 in
\citet{Hirata2004} to estimate this bias,
\begin{equation}
\frac{\delta\gamma}{\gamma}=({\cal R}^{-1}-1)\frac{\delta T^p}{T^p}\,.
\end{equation}
As in H04 we fix $\frac{\delta T^p}{T^p}$ to be 0.03.  Due to the fact
that the PSF reconstruction pipeline and the PSF size are fixed, $T$
varies very little. The estimated bias is $\pm 2.2$ per cent. That is
consistent with M05 at $\pm 2.1$ per cent for $r<21$ and $\pm 2.4$ per
cent for $r>21$. Our estimate is between these two values because we
calculate the bias using all the galaxies with $r$ band model
magnitude.

\subsubsection{PSF dilution bias}

The PSF blurs the image due to the convolution, which is a function of
resolution ${\cal R}$ and brightness distribution.  An empirical
formula of this bias from an ensemble of exponential and de
Vaucouleurs distributions is given in M05 as a function of the
fraction of exponential part and the fraction of de Vaucouleurs part,
\begin{equation}
\frac{\delta\gamma}{\gamma}\geq -0.014f_{exp}-0.035f_{deV}\,.
\end{equation}
Roughly, this value ranges from -2.8 to 3.9 percent. As both studies
use PHOTO PSF pipeline and because this bias is estimated in a model
dependent method, we directly use M05's estimation as shown in
Table. \ref{tab:tbl-2} (the * symbol indicates that we directly use
M05's results).

\subsubsection{Shear responsivity error}

The responsivity $\bar{R}$ is calculated from the variance of
ellipticity, indicating that this is related to the ellipticity
distribution. Once we use the cut ${\cal R}>1/3$, the distribution has
been changed and an error on $\bar{R}$ appears. It ranges from 0 to
1.7\% in M05.  Our bias estimation using Eq. 25 in H04 is 2.3\% with a
fixed $\delta e_{rms}=0.02$ as in H04.

\subsubsection{Noise rectification bias}

This noise, ranging from $-3.8$ to 0\%, is caused by the image noise
as described in HS03 (Eq.26 and Eq.27). The quantification of this
bias is,
\begin{equation}
\frac{\delta\gamma}{\gamma}\approx Kv^{-2}=
4(1-3\bar{R}_2^{-1}+\bar{R}_2^{-2}+2e_{rms}^2)v^{-2}\,,
\end{equation}
where $v$ is the signal-to-noise ratio of the detection over bands
$v^{-2}=\frac{2}{v_r^2+v_i^2}$. Our estimate of $K$ at ${\cal R}=1/3$
is 5.7 bigger than 5.3 in M05 and 5.1 in H04. So the lower limit of
this bias ($2\sigma$) in our sample is $-4.08$ per cent larger than
M05 and H04.

\subsection{Systematic tests}

In order to observationally estimate the systematics, three additional
tests were carried out: a redshift test, a random sample test and a 45
degree rotation test. Any systematics will cause a deviation from the
expected zero.

\subsubsection{Redshift test}

\begin{figure}
\centering
\includegraphics[width=7cm,height=9cm]{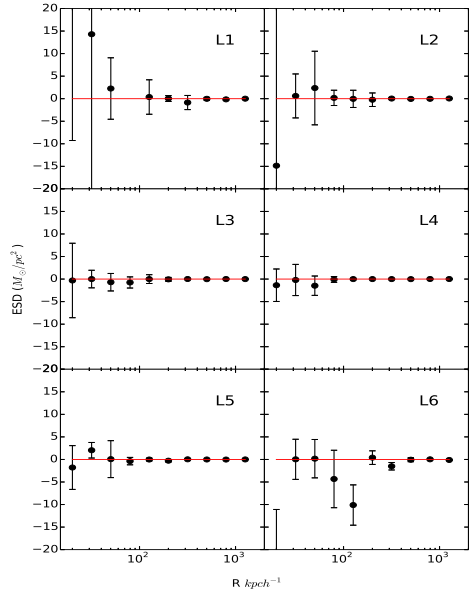}
\caption{Redshift systematic test. Shown in the plot are the ESDs
estimated using sources galaxies that are in front of the lens
galaxies. }
  \label{fig:test1}
\end{figure}

The redshift test is performed first. The lens-source separation used
for shear calculation is $z_l<z_s+0.1$. This criteria has been chosen
to avoid cases where the source galaxy may be located in front of the
lens galaxy. The value, $0.1$ is chosen on the basis that the typical
photometric redshift measurement error is $0.025$
\citep{Abazajian2009}.  If we use $z_l>z_s$, no signal is expected,
and non-zero value would be caused by unknown systematics.
Fig. \ref{fig:test1} shows this systematic test using our SDSS DR7
data. The consistency with zero shows that the systematics in our work
can be neglected in comparison to the null lensing signals.

\subsubsection{Random sample test}

\begin{figure}
\centering
\includegraphics[width=7cm,height=9cm]{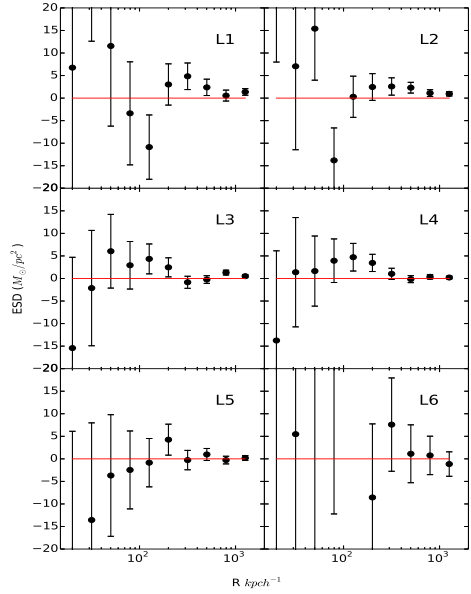}
\caption{Random sample test,  the ESDs
  estimated around random lens galaxies.}
  \label{fig:test2}
\end{figure}

For this test, we have used the random catalog constructed in
\citet{Yang2012}, which was used to calculate the two point
correlation function. This random sample includes all the
observational effects from SDSS DR7, i.e., the same luminosity
function, magnitude limit, redshift completeness and sky coverage due
to SDSS mask (MANGLE by \citet{Hamilton2004}). The total number of
random galaxies in the sample is 736,812, slightly larger than the
original sample we used. We binned the random sample into the same 6
luminosity ranges and measured the galaxy-galaxy lensing signals
around the random samples. Fig.\ref{fig:test2} shows the signals
obtained around the random samples, which are all consistent with null
signals within the one sigma uncertainties.

\subsubsection{45 degree rotation test}

Finally, we calculate the B mode signal using all the galaxies. As in
M05, we calculated the 45 rotated signals with 4 distance bins, i.e.,
$30 <R<100\kpch$, $100 <R<600\kpch$, $600<R<2000\kpch$ and $30
<R<2000\kpch$.  Again, this systematic is consistent with zero within
the one sigma error, as shown in the following table.

\begin{table}[h!]
\begin{center}
\caption{\label{tab:tbl-3} This table shows the results
   of 45 degree rotation tests as in M05. }
\begin{tabular}{ccc}
\hline
Radial range($\kpch$) & $\Delta\Sigma_{45}(hM_{\odot}{\rm pc}^{-2})$ & $\sigma_{45}$\\
\hline
$30<R<100$   & -0.46  & 1.42    \\
$100<R<600$  & 0.02   & 0.24    \\
$600<R<2000$ & -0.01  & 0.10    \\
$30<R<2000$  & -0.11  & 0.12    \\
\hline
\end{tabular}
\end{center}
\end{table}

\section{ The ESDs of lens galaxies}
\label{sec_ESD}

In this appendix we provide the ESD measurements for our SDSS DR7 lens
samples in different luminosity and stellar mass bins, in Tables
\ref{tab:ESD-L6} and \ref{tab:ESD-sm7}, respectively.  ESDs for galaxy
subsamples separated by color and by star formation rate, along with
{\it all} the relevant covariance matrixes are provided in electronic
files publicly available via the link
\url{http://gax.shao.ac.cn/wtluo/weak\_lensing/wl\_sdss\_dr7.tar}.

\begin{table}[h!]
\begin{center}
 \caption{\label{tab:ESD-L6} This table lists the ESDs of lens
   galaxies that are separated into different luminosity bins}.
\begin{tabular}{cccccccc}
\hline
$R (Mpc/h)$ & L1 & L2 & L3 & L4 & L5 & L6 \\
\hline
0.020 & $47.753 \pm 51.318$  & $22.722 \pm 31.109$   & $59.068 \pm 21.041$    & $85.001 \pm 17.596$  & $270.283 \pm 19.675$   & $395.426 \pm 112.376$  \\
0.032 & $50.609 \pm 29.971$  & $29.589 \pm 19.232$   & $28.174 \pm 13.808$    & $59.956 \pm 12.268$  & $58.845  \pm 15.307$   & $199.987 \pm 50.518 $  \\
0.050 & $27.104 \pm 18.650$  & $6.401  \pm 12.476$   & $35.059 \pm 8.693 $    & $33.181 \pm 7.948 $  & $54.852  \pm 12.084$   & $96.381  \pm 45.461 $  \\
0.080 & $0.760  \pm 11.775$  & $8.969  \pm 7.692 $   & $18.018 \pm 5.313 $    & $27.231 \pm 5.129 $  & $35.538  \pm 8.089 $   & $111.051 \pm 31.248 $  \\
0.126 & $6.225  \pm 7.465 $  & $8.183  \pm 5.040 $   & $9.380  \pm 3.491 $    & $12.053 \pm 3.055 $  & $21.476  \pm 4.981 $   & $105.765 \pm 22.670 $  \\
0.200 & $7.674  \pm 4.772 $  & $6.517  \pm 3.260 $   & $12.707 \pm 2.162 $    & $11.417 \pm 1.920 $  & $17.160  \pm 3.042 $   & $54.225  \pm 13.056 $  \\
0.317 & $3.839  \pm 3.063 $  & $3.512  \pm 1.990 $   & $4.422  \pm 1.381 $    & $8.229  \pm 1.269 $  & $14.231  \pm 1.957 $   & $38.240  \pm 8.547  $  \\
0.502 & $1.463  \pm 1.929 $  & $5.645  \pm 1.282 $   & $4.528  \pm 0.860 $    & $4.891  \pm 0.783 $  & $11.296  \pm 1.183 $   & $18.851  \pm 4.931  $  \\
0.796 & $2.557  \pm 1.266 $  & $4.166  \pm 0.797 $   & $4.106  \pm 0.551 $    & $3.616  \pm 0.526 $  & $5.322   \pm 0.758 $   & $17.179  \pm 3.135  $  \\
1.261 & $4.227  \pm 0.773 $  & $3.990  \pm 0.561 $   & $3.191  \pm 0.358 $    & $2.979  \pm 0.333 $  & $4.735   \pm 0.499 $   & $10.577  \pm 1.931  $  \\
\hline
\end{tabular}
\end{center}
\end{table}
 
\begin{table}[h!]
\begin{center}
 \caption{\label{tab:ESD-sm7} This table lists the ESDs of lens
   galaxies that are separated into different stellar mass bins}.
\begin{tabular}{ccccccccc}
\hline
$R (Mpc/h)$ & sm1 & sm2 & sm3 & sm4 & sm5 & sm6 & sm7 \\
\hline
0.020 & $-4.333 \pm 37.732$  & $32.952 \pm 31.256$   & $82.383 \pm 21.056$    & $182.637 \pm 21.271$  & $335.980 \pm 33.494$   & $336.404 \pm 30.651$ & $115.285 \pm 99.517$ \\
0.032 & $36.117 \pm 25.027$  & $-14.994\pm 19.186$   & $41.799 \pm 14.208$    & $91.604  \pm 15.367$  & $43.176  \pm 21.400$   & $31.518 \pm 22.789$ & $256.227 \pm 58.129 $ \\
0.050 & $16.766 \pm 15.203$  & $30.626 \pm 12.198$   & $34.166 \pm 9.904 $    & $44.174  \pm 11.545$  & $73.646  \pm 18.962$   & $66.753 \pm 18.781$ & $123.575 \pm 45.973 $ \\
0.080 & $21.305 \pm 9.069 $  & $10.033 \pm 7.549 $   & $20.878 \pm 5.701 $    & $33.090  \pm 7.798 $  & $52.717  \pm 12.550$   & $49.856 \pm 13.153$ & $113.419 \pm 35.431 $ \\
0.126 & $10.617 \pm 5.837 $  & $13.010 \pm 4.850 $   & $5.807  \pm 3.915 $    & $23.703  \pm 4.843 $  & $38.787  \pm 8.170 $   & $40.202 \pm 7.755 $ & $114.243 \pm 25.088 $ \\
0.200 & $7.286  \pm 3.841 $  & $12.030 \pm 3.186 $   & $14.715 \pm 2.401 $    & $12.398  \pm 2.847 $  & $24.732  \pm 4.965 $   & $23.755 \pm 4.770 $ & $65.843 \pm 15.455 $ \\
0.317 & $3.062  \pm 2.413 $  & $3.039  \pm 1.921 $   & $9.233  \pm 1.482 $    & $9.110   \pm 1.803 $  & $22.772  \pm 3.356 $   & $22.250 \pm 3.147 $ & $43.172 \pm 9.255 $ \\
0.502 & $3.836  \pm 1.496 $  & $3.722  \pm 1.273 $   & $5.069  \pm 0.965 $    & $9.217   \pm 1.174 $  & $15.316  \pm 2.050 $   & $14.874 \pm 1.904 $ & $22.356 \pm 5.645 $ \\
0.796 & $2.573  \pm 0.975 $  & $3.312  \pm 0.817 $   & $3.807  \pm 0.605 $    & $4.872   \pm 0.726 $  & $7.290   \pm 1.218 $   & $7.172 \pm  1.273 $ & $21.094 \pm 3.776 $ \\
1.262 & $2.929  \pm 0.570 $  & $3.503  \pm 0.497 $   & $3.155  \pm 0.383 $    & $4.023   \pm 0.480 $  & $5.760   \pm 0.786 $   & $5.703 \pm 0.749 $ & $12.316  \pm 2.337 $ \\
\hline
\end{tabular}
\end{center}
\end{table}

%

%\bibliography{ms}

\end{document}